\documentstyle[epsf,epsfig]{mn2e}

\newcommand{\kms} {km s$^{-1}$\ }
\newcommand{\msol} {M$_{\odot}$}
\newcommand{\zsol} {Z$_{\odot}$}
\newcommand{\rsol} {R$_{\odot}$}
\def\lesssim{\mathrel{\hbox{\rlap{\hbox{\lower4pt\hbox{$\sim$}}}\hbox{$<$}}}}
\def\gtrsim{\mathrel{\hbox{\rlap{\hbox{\lower4pt\hbox{$\sim$}}}\hbox{$>$}}}}
\title[The progenitor star of SN~2001du]{Mass limits for the progenitor star of supernova 2001du and other type II-P supernovae}
\author[Smartt et al.]{S. J. Smartt$^{1}$, J. R. Maund$^{1}$, G. F. Gilmore$^{1}$,  C. A. Tout$^{1}$, D. Kilkenny$^{2}$, S. Benetti$^{3}$,\\
$^{1}$Institute of Astronomy, University of Cambridge, Madingley Road, Cambridge CB3 0HA, England\\
$^{2}$South African Astronomical Observatory, PO Box 9, Observatory 7935, South Africa\\
$^{3}$INAF - Osservatorio Astronomico di Padova, vicolo dell'Osservatorio 5, I-35122 Padova, Italy\\
}
\begin{document}

\date{Accepted. Received 2003 January;  in original form 2003 January}

\pagerange{\pageref{firstpage}--\pageref{lastpage}} \pubyear{2002}

\maketitle

\label{firstpage}

\begin{abstract}

The supernova SN~2001du was discovered in the galaxy NGC1365 at a
distance of $19\pm2$\,Mpc, and is a core-collapse event of type
II-P. Images of this galaxy, of moderate depth, have been taken with
the Hubble Space Telescope approximately 6.6 years before discovery and
include the supernova position on the WFPC2 field of view. We have
observed the supernova with the WFPC2 to allow accurate differential
astrometry of SN~2001du on the pre-explosion frames. As a
core-collapse event it is expected that the progenitor was a massive,
luminous star. There is a marginal detection (3$\sigma$) of a 
source close to the supernova position on the prediscovery $V-$band 
frame, however it is not precisely coincident and we do
not believe it to be a robust detection of a point source. We conclude
that there is no stellar progenitor at the
supernova position and derive sensitivity limits of the prediscovery images
which provide an upper mass limit for the progenitor star. We
estimate that the progenitor had a mass of less than 15\msol. We
revisit two other nearby SNe II-P which have high quality
pre-explosion images, and refine the upper mass limits for the
progenitor stars. Using a new distance determination for SN~1999gi
from the expanding photosphere method,  
we revise the upper mass limit to 12\msol. We present new HST images
of the site of SN~1999em, which validate the use of lower spatial
resolution ground-based images in the progenitor studies and 
use a new Cepheid distance to the galaxy to measure 
an upper mass limit of  15\msol\ for that progenitor. Finally
we compile all the direct information available for 
the progenitors of eight nearby core-collapse supernovae and compare
their mass estimates. These are compared with the latest stellar evolutionary
models of pre-supernova evolution which have attempted to relate
metallicity and mass to the supernovae type. 
Although this is statistically limited at present, reasonable agreement 
is already found for the lower mass events (generally the II-P), but 
some discrepancies appear at higher masses. 

\end{abstract}

\begin{keywords}
galaxies: individual (NGC1365) -- stars: evolution -- supernovae: general --
supernovae: individual (SN~2001du, SN~1999em, SN~1999gi)
\end{keywords}

\section{Introduction}
The end point of the lives of all stars with initial mass greater
than approximately 8\msol\ is thought to be a core-collapse supernova. 
Supernovae (SNe) explosions which arise from massive
stars have been a major driver in 
the chemical evolution of the Universe, are  important
in shaping the interstellar medium dynamics in gas-rich galaxies
and are fundamental in the study of the origin of the chemical 
elements in the Universe. The exact evolutionary states of
stars before they undergo core-collapse is of great interest in the
fields of massive star evolution and the physics of supernovae evolution. 

Supernovae are classified according to the atomic 
lines observed in the early-time optical spectra. 
Briefly, the presence of H\,{\sc
i} optical lines indicate a type~II classification (SNe II), while those
that show no hydrogen at all are type\,I \cite{fili97}. The SNe~Ia
are thought to arise through thermonuclear explosions in white dwarf
binary systems \cite{branch95}. Given the occurrence of these
supernovae in both old and intermediate age stellar populations
(e.g. in galaxies of type SO and earlier), the white dwarf binary
scenario is consistent with the progenitors being low$-$intermediate
mass stars. However very little other details are known about the
stellar systems that produce SNe~Ia.  All other supernovae including
the type Ib/Ic (SNe Ib/Ic) and all the subtypes of SNe II are thought
to be due to core-collapse during the deaths of massive stars. The
subtype SNe II-P show prominent, broad H\,{\sc i} lines with P-Cygni
type profiles in their optical spectra, indicating that the progenitor
retained a substantial hydrogen envelope prior to explosion 
e.g. see Woosley \& Weaver \shortcite{ww86}. They also display a plateau phase
in their lightcurve during which the supernovae remain at approximately
constant brightness for a period of between 30 and 100 days, which is
also  interpreted as evidence of an extended hydrogen atmosphere
in the progenitor star, such as red supergiant 
(Chevalier 1976, 
Woosley \& Weaver 1986). This plateau phase is most obvious in the
$V-$band, but can less pronounced or non-existant at other visual
and NIR wavelengths.
The type-IIn supernovae show 
strong hydrogen features which are in emission with broad wings and
very narrow cores. These features are thought to arise from
interaction with dense circumstellar material which may have been
ejected from the progenitor star through stellar winds or more
dramatic mass-loss events e.g. see Fassia et al. \shortcite{fass01}, 
Pastorello et al. \shortcite{pass02}, 
Lentz et al. \shortcite{lentz01}
and references therein.  The SNe~Ib/Ic also do not show any obvious
signs of hydrogen in their spectra, although SNe Ib display pronounced
He\,{\sc i} absorption. However recently Branch \shortcite{branch02} has
suggested that at least some Ib events do show some amounts of
hydrogen, and some Ic events do show helium. It is likely there is a
continuum of possibilities rather than the well defined sub-classes
traditionally employed.

There is very strong evidence that the SNe II and Ib/Ic are associated
with the deaths of massive stars as they are never seen in elliptical
galaxies, only rarely in S0 types \cite{vanden02} and they often appear to be
associated with sites of recent starformation such as H\,{\sc ii}
regions and OB associations in spiral and irregular galaxies
\cite{vandyk96}. However the desired direct evidence
of detecting stars before they exploded has only been possible for a
small number of objects. The two most unambiguous detections of a
progenitor object of confirmed supernovae explosions are for SNe 1987A
and 1993J. The progenitor to SN~1987A in the LMC \cite{white87} was
the blue supergiant Sk$-69^{\circ}202$ of spectral type B3\,Ia
\cite{wal89}.  The closest supernova to the Milky Way since then was
SN~1993J in M81 (3.63\,Mpc), which was a type\,IIb event and ground
based $UBVRI$ photometry of the SN site before explosion was presented
by Aldering et al. \shortcite{alder94}. 
The photometry of the progenitor candidate was
best fit with a composite spectral energy distribution of a K0\,Ia
star and some excess $UB$ band flux either from unresolved OB
association contamination or a hot companion. Neither the progenitor
of SN~1987A nor that of SN~1993J is consistent with the canonical
stellar evolution picture, where core carbon burning finishes and
core-collapse occurs relatively soon afterwards ($\sim10^3-10^4$\,yrs)
while the massive star is an M-supergiant. Claims have been made on
the detection of the progenitors of three other supernovae events
SNe 1961V, 1978K and 1997bs, however for reasons we discuss in
Section\,\ref{discuss_all} it is  uncertain whether these were
actual supernovae explosions. 

The only way to further add to our knowledge of the core-collapse
process and its relation to massive stellar evolution is to identify
more progenitor objects. There is now a wealth of data in the Hubble
Space Telescope (HST) archive, as well as high quality ground-based
images from various well maintained archive facilities (e.g. ESO, ING,
CFHT). Thus there are reasonable chances of having available an image
of a pre-supernova site which has sufficient resolution to identify
individual massive stars. Clearly there is a distance limit within
which images are useful from depth and resolution constraints.  This
is roughly 10\,Mpc for a deep, good quality (natural seeing of
typically $0.7''$) image from a ground-based 2-8m telescope, and
roughly 20\,Mpc for typical exposure length HST frames. Recently
Smartt et al.  \shortcite{sma01a,sma02a,sma02b} have used both HST and
ground based images of pre-explosion supernovae sites to try to detect
progenitor stars of SNe II-P 1999gi and 1999em and the type Ic
SN~2002ap. These nearby events have high quality prediscovery archive
images available, but no progenitor star is actually identified in any
of these cases.  These papers discuss the luminosity limits which can
be placed on the exploding star from the sensitivity measurements of
the images, and quite restrictive mass-limits can be determined (see
Section\,\ref{discuss_all} for a full comparison and discussion).  In
a similar vein, Van Dyk et al. \shortcite{vandyk03a} have attempted to
identify the progenitors of 16 type II and type Ib/s SNe on HST
prediscovery images. However 9 of the objects in this lengthy paper
are further than $\sim$40\,Mpc, with another reddened by some $5-6$
magnitudes of visual extinction, hence there is little or no chance of
identifying a progenitor star for these events.  The other six are
subject to the detailed HST follow-up programme of Smartt et al. (GO
9353) which will allow more accurate astrometry and possible 
source detection. The Van Dyk et al. study has attempted absolute
astrometry of WFPC2 archive images as a method of determining the 
supernovae positions on the prediscovery frames. This method has
not proved accurate enough to position the events to 
the precision necessary to assess if a progenitor star can be detected.
The error boxes the define are generally too large to provide
useful constraints. They do sugges that there are possible 
progenitor candidates for four supernovae, although these will require 
more detailed follow-up to confirm or discount the claims. 

In this paper we present a search for a progenitor of the 
core-collapse supernova SN~2001du. This was a type II-P event in the
galaxy NGC1365 at a distance of $(m-M)=31.39\pm0.26$ (Ferrarese et al
2000; corresponding to $19\pm2$\,Mpc) and the supernova position was
observed by HST in 1995. The supernova was discovered by Evans
\shortcite{evans01} on 2001 August 24.7 UT at a visual magnitude of
about 14.0. It was classified by Smartt et al. \shortcite{sma01c} on
2001 September 2.05 UT, and Wang et al. \shortcite{wang01} on August
30, as a type II-P supernova due to the presence of strong H\,{\sc i}
P-Cygni profiles in the H$\beta$ and H$\alpha$ transitions. Wang et
al. also obtained spectropolarimetry of SN~2001du and reported the
supernova to be polarized at a level of 0.29\% with a flat spectrum
characteristic of electron scattering, indicative of a fairly spherical
explosion. This supernova has very similar optical spectra to the other 
type II-P events SNe 1999em and 1999gi for which we determined
upper mass limits for the progenitor star.
Images of the SN~2001du host galaxy 
are available in the HST archive allowing another detailed
search for a precursor object. In this paper we  
show that there is a marginal detection of a faint source
close to the supernova position. But it is neither a 
robust detection of a single point source nor precisely 
coincident with the supernova.
We argue that there is no detection of 
a progenitor, and hence determine mass
and luminosity limits of the exploding star and 
discuss the implications of this finding
alongside all the available, direct information we have on
progenitors of core-collapse supernovae. 

\begin{table*}
\caption{Details of the HST WFPC2 archival and GO 
observational data for SN~2001du. The dataset names come from the 
the ST-ECF Science archive (http://archive.eso.org)
and are the associations of shorter, aligned exposures.
The location 
column indicates where SN~2001du occurred on the WFPC2 FOV.  
The PC1 has a plate scale of $0.0455''$ per pixel, 
WF3 has $0.0996''$ per pixel.}
\label{obsdata}
\begin{tabular}{lllcc}
\hline\hline 
Dataset &  Date & Filter & Exposure Time  & Location  \\ \hline
U2KV010AB    & 1995 January 15th  & F160BW & 330s   & WF3\\
U2KV010AB    & 1995 January 15th  & F336W  & 330s   & WF3\\
U2KV010JB    & 1995 January 15th  & F555W  & 330s   & WF3   \\
U2KV010SB    & 1995 January 15th  & F814W  & 330s   & WF3   \\
\\
U6BR030JB    &  2001 November 26th & F336W  &  560s  & PC1    \\
U6BR030NB    &  2001 November 26th & F439W  &  400s  & PC1    \\
U6BR0303B    &  2001 November 26th & F555W  &  640s  & PC1    \\
U6BR030LB    &  2001 November 26th & F675W  &  400s  & PC1    \\
U6BR030BB    &  2001 November 26th & F814W  &  640s  & PC1    \\
\hline                                            
\end{tabular}   
\end{table*}

\section{Observational data and analysis }

\subsection{Astrometry and detection limits of WFPC2 images}
\label{data}
The galaxy NGC1365 was observed on four epochs with the WFPC2 on board
HST before the supernova 2001du was discovered. However on only one of
these occasions does the supernova position actually fall on any of
the WFPC2 chips. The galaxy was a target for the HST key project on
the extragalactic distance scale, which was designed to accurately
resolve and photometrically measure
Cepheid variables in this spiral. Unfortunately
these deep exposures (totalling 18.5hrs in F555W and 6hrs in F814W) do
not cover the position of the progenitor of SN~2001du. The observations
which do include the supernova position were taken in Cycle\,4 in
GO5222 (PI: J. Trauger), and the details are listed in
Table\,\ref{obsdata}. These data were located and retrieved 
from the ST-ECF Science archive using the {\sc astrovirtel} initiative. 
The approximate position of the supernova on
these prediscovery images can be estimated \cite{vandyk03a}, however one
is limited by the absolute astrometric resolution of the WFPC2 frames
(usually of the order $1-2''$) and the error on the SN position
($\sim0.5-1''$).  As this could amount to $\sim$30 pixels on the WF
chips of WFPC2, it is essential that more accurate astrometry is
achieved to determine the SN position on the prediscovery frames. As
discussed in Smartt et al. (2001a), by far the best
method is to take repeat WFPC2 observations of the supernova some
months after explosion and use the surrounding stars in the post and
pre-explosion images to determine precise {\em differential}
astrometry.  The crowded host region of SN~1999gi shown by
Smartt et al. (2001a) and Li et al. (2002)
illustrates that an RMS error of even $0.2''$ in positioning the
supernova would not have allowed two nearby, resolved point sources to
be have been ruled out as the progenitor objects. Hence re-observing
the SN at the resolution of HST (with either WFPC2 or ACS) is
essential to accurately determine the SN position.

\begin{figure*}
\caption{Panels (a)-(c) show images of the F336W, F555W and F814W
filters taken 6.6 yrs before explosion transformed to the pixel
coordinates of the post-explosion (d). The images (a)-(c) are from
the WF3 chip ($0.1''$ pixels), and (d) is better sampled with 
the $0.05''$ pixels of the PC1. The position of the SN is marked as a 
cross in the prediscovery images, and the blank triangular regions are
due to the edge of the WF3 chip after transformation. The fields of view
of these panels are $20''$, corresponding to a distance of 1.8\,kpc
at the distance of NGC1365}
\label{prepostimage}
\end{figure*}

\begin{figure*}
\epsfig{file=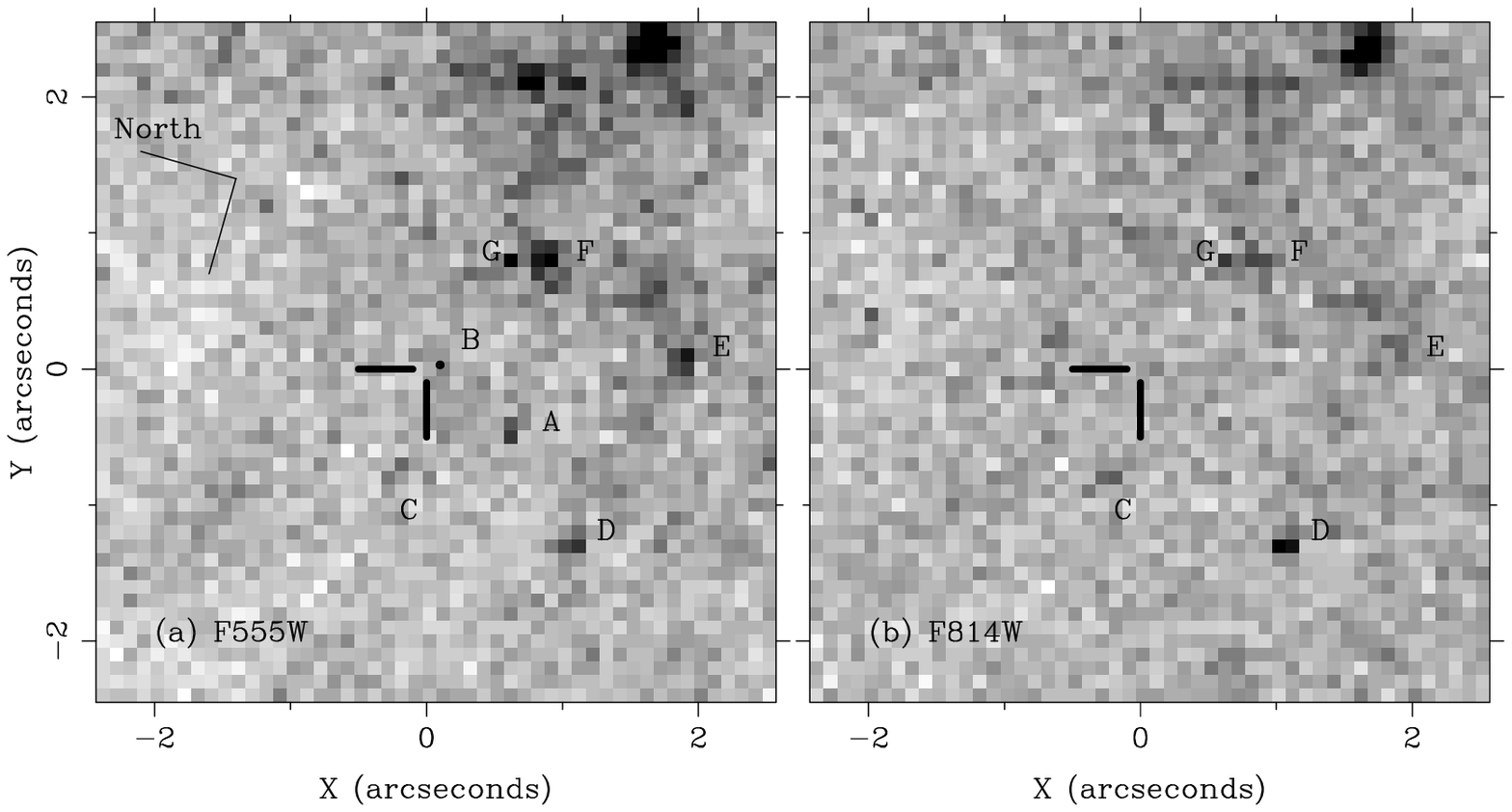,width=16cm}
\caption{A close up of the original F555W and F814W pre-explosion
images before the transformation described in Section\,2.1. The
position of the SN is marked at position (0,0) with the orthogonal
lines. The width of these lines represent the 
combined error in its position and the transformation
RMS fit errors ($0.14''$). We keep the same labeling as 
Van Dyk et al. (2003a) for stars A-B-C, and extend it to the 
other objects detected in the field. The possible object (star
B) detected at 3$\sigma$ above the background noise is labeled, 
and its position is marked with a solid dot. The size of this dot
represents the probable centroid error on the position of the
source, and within the errors it is not coincident with the 
supernova. Star B is difficult to identify visually, and is 
more likely a variation in the complex background rather than a 
true point source. It is not detected in the F814W filter. 
The size of the image panels are $5''$,
corresponding to a distance of of 460\,pc at the distance of
NGC1365. The directions of north and east are marked. }
\label{prepostzoom}
\end{figure*}

The supernova was observed with WFPC2 on 2001  November 26 through
the program GO9041, and the exposure times and filter details are
listed in Table\,\ref{obsdata}. These are total exposure times, and as
the SN was still quite bright at this epoch, the F336W, F555W and
F814W were broken down into sub-exposures to allow photometry of a
non-saturated SN point-spread-function (PSF) as well as the fainter
background stars. Both the F555W and F814W datasets are composed of a
40s and $3\times$200s exposures, while the F336W is composed of 100s
and 460s exposures (with CR splits for cosmic ray removal).
Photometry of stars visible in both these long and short exposures
showed excellent alignment to better than 0.1\,pixel in each dimension. 
Hence the centroid position of the supernova can be measured in 
each of the short frames, and assumed to be appropriate for the
longer frames. Seventeen bright ($V_{555}\simeq20.5-23.5$), 
 single stars were identified
in a $500\times500$ pixel region around the SN position in the 
F555W post-explosion 600s exposure and their positions measured 
with aperture photometry within the IRAF DAOPHOT package. The positions of 
these same 17 stars were identified in the prediscovery frames and 
a geometric solution mapping these positions to the post-explosion frame
was calculated. The spatial transformation fitted a linear shift, 
magnification factor and a rotation, giving an RMS to the fit of 
$0.01''$ in each of $x$ and $y$. Eleven stars were used in the F814W
transformation, with similar RMS errors. In Fig.\,\ref{prepostimage}
we show the position of the supernova in the 3 colour prediscovery 
images. The total exposure time available for all the stacked 
images at this position is 330s in each filter. However
this is made up of 3$\times$100s and 3$\times$10s individual frames, 
Given the significant read-out noise of the WFPC2 chips (7e$^{-}$ per frame)
compared with the sky background in a stellar aperture, the 
3$\times$10s frames were ignored, and only the 300s stacked images
were used. We note that Van Dyk et al. \shortcite{vandyk03a}
quote an exposure time of 100s total for these images, and presume
this is a typographical error. 

Very close to the position of the supernova on the F555W image there is a
possible detection of a faint source which is close to the detection
limit of the image. Assuming Poisson statistics, a total CCD readout
noise of 12e$^{-}$ (from the three individual stacked frames), and a
gain of 14e$^{-}$/ADU, this object is detected with a 3.0$\sigma$
significance above the background noise. The magnitude of the object
estimated by placing a 2-pixel radius aperture at the position 
is $V_{\rm 555}\simeq25.0\pm0.4$. This was obtained 
using the IRAF DAOPHOT package, and applying the appropriate
charge-transfer-efficiency \cite{whit99}
and aperture corrections \cite{holtz95a}. A
model point-spread-function was calculated using the Tiny Tim software
\cite{kh99}, and PSF fitting photometry was also carried out on the
source using the DAOPHOT task ALLSTAR. A consistent magnitude of
$V_{\rm 555}\simeq25.2\pm0.4$ is measured. However it is a marginal
detection, the object is not clearly identified by eye on the frames
and is not a convincing point source. Placing a 2 pixel aperture
centred on various pixels in the vicinity 
which are individually more than 2$\sigma$ above their neighbours
will frequently give a measured magnitude of between 25.0$-$26.2. 
The source appears somewhat extended, and has two peaks of intensity
separated by roughly two pixels. After PSF subtraction one of these
pixel peaks always remains, which suggests to us that this is not a
robust detection of a point source. The source is not detected 
in the F814W filter. Furthermore, the
supernova position is not exactly coincident with the object within the
estimated errors of our methods. We estimate the error radius on the
precursor object position to be 0.3 pixels, which is the maximum
difference between positional determinations using four methods
i.e. intensity weighted means in each dimension, a Gaussian fit, a
triangular weighting function, and the PSF fit.  
However we find the object
1.03\,pixels from the supernova centroid. The error in the
supernova centroid is 0.14\,pixels, 
and transformation
error as discussed above is 0.14 pixels RMS, giving a total error circle
of radius $r_{\rm err}=0.28$\,pixels. Hence considering the errors
in the analysis,  the supernova position is not precisely coincident with this
possible source (see Fig.\,\ref{prepostzoom}). 

We have followed the labelling of stars in the region
from Van Dyk et al. \shortcite{vandyk03a}, who identify three
objects within a $0.9''$ error circle of their estimated supernova
position, but the only unambiguous detection we confirm is that of 
star A. The object we marginally detect close to the supernova
position is the Van Dyk et al. 
star B, and we keep the nomenclature for clarity. 
Their star C is also a marginal detection, but as it is some way
from the supernova position, we do not consider it any further. 
These images are available in the HST archive, and we
quote our determination of the position of SN~2001du (at pixel
(106.80, 675.96), with errors as listed above) on
chip WF3, to allow our results to be independently verified if
necessary. There doesn't appear to be a significant discrepancy 
between our results and those of Van Dyk et al., as the latter 
have used the HSTphot photometry routines of 
Dolphin \shortcite{dolph00a,dolph00b} with a 3$\sigma$ threshold. 
We do repeat a 3$\sigma$ detection, but for the reasons outlined
above do not interpret it as a certain detection of a stellar 
progenitor. Van Dyk et al. measure a magnitude of 25.02$\pm$0.3,
which is in good agreement with our PSF magnitude given the errors, 
the faintness of the source and the different methods employed. 

We 
determined the detection limits of the data in the vicinity of the
SN position using two methods. The first is just a simple calculation
of the limiting sensitivities based on Poisson statistics assuming the
CCD parameters given above. For a 2 pixel aperture, these are 
$V_{\rm 555}(3\sigma)=25.3$  and $V_{\rm 555}(5\sigma)=24.8$. 
Correcting for the CTE and applying an aperture correction 
would reduce these to $V_{\rm 555}(3\sigma)=25.0$ and 
$V_{\rm 555}(5\sigma)=24.5$. 
To validate this result, we performed simulations of a
synthetic star close to the supernovae position, (at a point 
with very similar sky values to those surrounding the possible star
B). Stellar photometry using PSF fitting procedures within the IRAF
package DAOPHOT was carried out, and all the stellar sources within
the region of the supernova were subtracted from the image.  The Tiny
Tim package \cite{kh99} was used to construct a suitable PSF for this
fitting procedure. Model PSFs of varying magnitudes between
24.4$-$25.6 were made, again using Tiny Tim, and these were added to
the ``flat'' image. A synthetic star at the magnitude of 
star B ($V_{\rm 555}\simeq$25.0) is not recovered with convincing
significance. Placing an aperture at the stellar position 
gives a measurement of $V_{\rm 555}\simeq25.7\pm0.9$. 
At the theoretical 5$\sigma$ limit of $V_{\rm 555}\simeq$24.5, 
the object can be visually identified and a magnitude of 
24.8$\pm$0.5 is determined.

We recognise that there
is some statistical validity in the 3$\sigma$ detection of star B
by Van Dyk et al. (2003a) and ourselves. 
However we cannot confirm, with any certainty,
an unambiguous detection of a point source. It looks more likely
to us that this object is a fluctuation of a 
fairly complex non-stellar background, 
which could loosely be interpreted as a point source, and that the 
progenitor is below (and possibly just below) the sensitivity limit. 
In the rest of this paper we assume that we do not have a secure 
identification of the progenitor, and base our results on the 
robust 5$\sigma$ upper limit to its $V_{\rm 555}$ magnitude. 
We agree with Van Dyk et al. that there is no sign of this object
in the other three filters, and 
exactly the same method for artificial star additions was 
applied to F814W and F336W, resulting in similar 5$\sigma$
detections at $I_{\rm 814}$=24.8 and $U_{\rm 336}$=21.6. The 
image in the UV filter also shows nothing to a limiting magnitude 
of $m_{\rm F160BW}=19.8$, and as this isn't particularly useful 
for placing constraints on the progenitor or its environment it 
is not discussed any further. 

\begin{figure*}
\epsfig{file=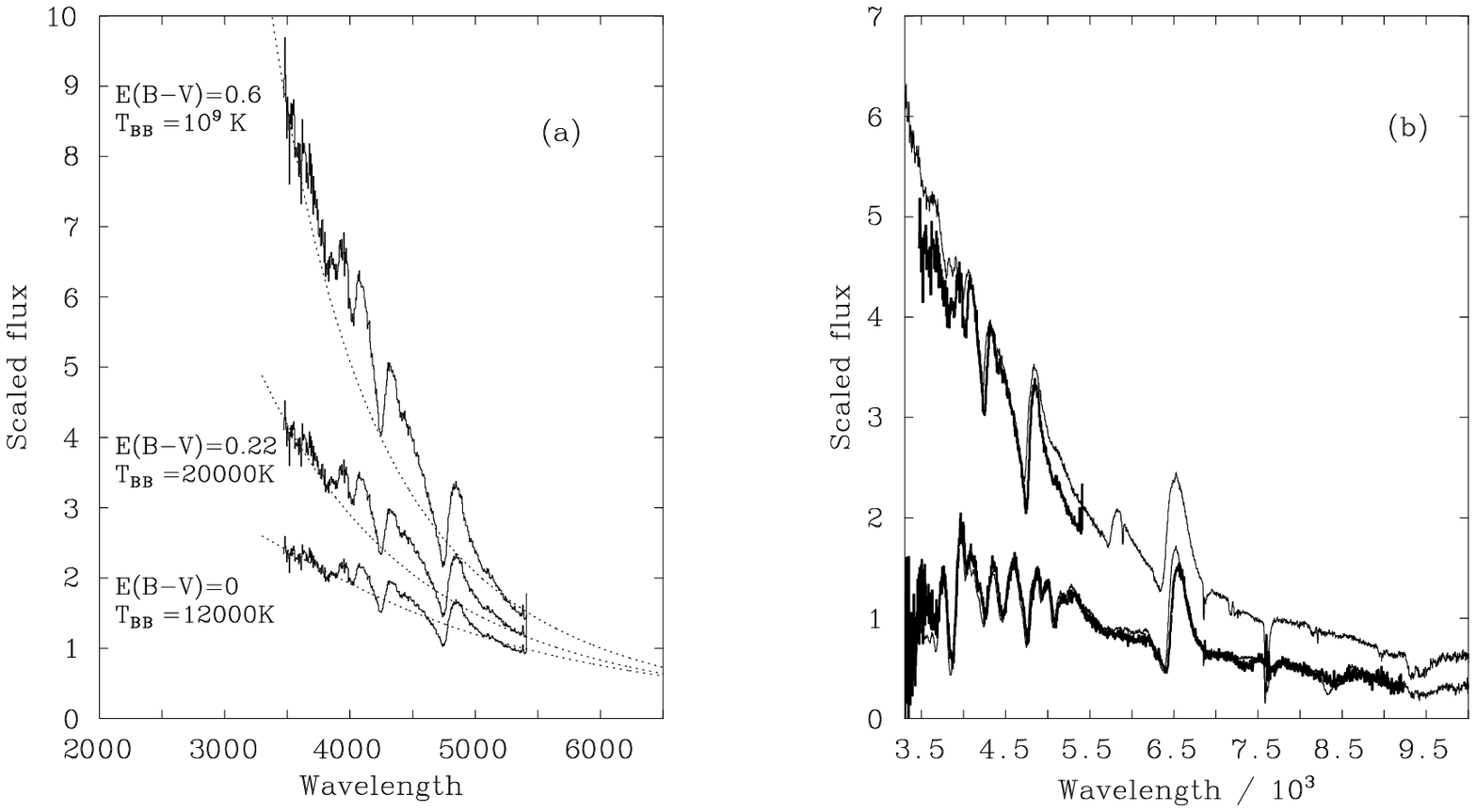,width=18cm}
\caption{{\bf (a):} The dereddened early time spectrum of SN~2001du taken 
on Sept. 2.05 is compared with black-body continuum fluxes
of various temperatures (dotted lines). Comparisons with 
an unphysically large temperature of $T_{\rm BB}=10^9$K effectively
set a hard upper limit to the reddening. If the spectrum is 
dereddened by more than $E(B-V)>$0.6, then the continuum cannot be
matched by any temperature. For comparison the more likely
temperatures of the photosphere of $T_{\rm BB}=12000$K and 20000K
(see Baron et al. 2000), are shown with the reddening required
to match. 
{\bf (b):} Comparisons of the optical spectra of SN~2001du 
(heavy line) with SN~1999em (lighter line -- from Hamuy et al. 2001).
The upper spectra are SN~2001du taken on  Sept. 2.05
and SN~1999em taken +7 days after explosion. 
The continuum of SN~2001du is very blue similar to SN~1999em, and indicates
the reddening of the two cannot be very different. The difference in 
the slopes is likely due to the difference of around 
4 days in the epochs. 
The lower spectra are SN~1999em taken on day +18 after explosion,
and SN~2001du taken
on 2001 September 9.35 UT. The spectra are almost identical (after
a linear scaling), indicating that SN~2001du is most probably also 
around +18 days, and the reddening must be similar to that for 
SN~1999em, for which a value of $E(B-V)=0.1\pm0.05$ was measured.}
\label{2001du_1999emspec}
\end{figure*}

\subsection{Photometry of the stellar population around SN~2001du}
Photometry in the three filter images F555W, F675W and F814W taken
after explosion on 26 November 2001 was carried out in order to help
determine a reddening towards the surrounding stellar
populations. These images were chosen over the prediscovery frames as
they are deeper and have better spatial resolution (as SN~2001du is
on the PC1 chip). Photometry using PSF fitting
techniques was carried out on the stars on the PC1 chip only, as the
reddening towards other parts of the galaxy contained in the rest of
the WFPC2 chips is not applicable to our present study. The on-the-fly
recalibrated images were taken from the archive and the IRAF version
of DAOPHOT was used to do the photometry.  Aperture photometry was
first carried out with a 4.4\,pixel radial aperture. Model PSFs were
calculated using the Tiny Tim package \cite{kh99} in each filter, and
these were used in the DAOPHOT task ALLSTAR. The zero-points from the
WFPC2 headers were applied, and several corrections where
included. Firstly the charge transfer efficiency (CTE) correction was
determined \cite{whit99}, then the magnitudes were corrected for the
4.4\,pixel aperture using the encircled energy tables of Holtzmann et
al. \shortcite{holtz95a}. Finally a colour transformation was
calculated to determine magnitudes of the stars in the Johnson $VRI$
bands from the WFPC2 flight system \cite{holtz95b}. Results from the
photometry are discussed below in Section\,\ref{discuss_phot}.

\subsection{Spectroscopy of SN~2001du}

Optical spectra of SN~2001du were taken on two epochs at the South
African Astronomical Observatory (SAAO), and at the European Southern
Observatory. The first spectrum was taken on 2001 September 2.05 at
the SAAO 1.9m using the Cassegrain Grating Spectrograph, with the
No. 6 grating (600 lmm$^{-1}$) and SITe detector providing a coverage
of $3500-5400$\AA. The spectral dispersion was 1.1\AA\,pix$^{-1}$, and
the resolution measured from the width of the arclines was
approximately $3$\AA.  Our second spectrum was taken at
ESO-La Silla on the Danish
1.54m telescope on 2001 September 9.35. The DFOSC spectrometer
was employed with grism No. 4 and the 4k$\times$2k EEV CCD which
provided a dispersion of 2.8 \AA\,pix$^{-1}$, and useful coverage
between $3300-9200$\AA.  The spectra were reduced using standard
techniques within IRAF. The bias level was subtracted from the CCD
images, before flat-fielding, extraction and wavelength calibration
from arc line maps. A photometric standard star was observed on each
occasion for relative flux calibration.  Both of the reduced spectra
are presented in Fig.\,\ref{2001du_1999emspec}.

Smartt et al. \shortcite{sma01c} 
and Wang et al. \shortcite{wang01}
classified this as a type II-P supernova
from the appearance of the P-cygni like H\,{\i} Balmer lines, 
particularly H$\beta$. The unpublished lightcurve of 
Suntzeff \& Krisciunas as shown in Van Dyk et al. 
\shortcite{vandyk03b} shows it to have had a plateau phase of approximately
100 days. The spectrum and lightcurve is very similar to the well studied
SN~1999em which was a classic II-P event with a 95 day
plateau \cite{ham01,leon02b,elm02}, and further discussion of the 
similarities are discussed below in Section\,\ref{discuss_spec}.

\section{Theoretical stellar evolutionary models}
\label{stellar_models}

We have calculated a number of stellar evolutionary models to compare
with the luminosity constraints from the prediscovery images. Models
were constructed at every integer mass between 9$-$20\msol\ which is
the most interesting region probed by the observational data, plus
several higher masses at 25, 40 and 60\msol. As described in Smartt et
al.  \shortcite{sma02a}, the models come from the most recent version
of the Eggleton evolution program
\cite{eggleton1971,eggleton1972,eggleton1973}. The equation of state,
which includes molecular hydrogen, pressure ionization and coulomb
interactions, is discussed by Pols et al. \shortcite{pols95}.  The
initial composition is taken to be uniform with a hydrogen abundance
$X = 0.7$, helium $Y = 0.28$ and metals $Z = 0.02$ with the meteoritic
mixture determined by Anders \& Grevesse \shortcite{anders1989}.
Hydrogen burning is allowed by the pp chain and the CNO cycles.
Helium burning is explicitly included in the triple $\alpha$ reactions
and reactions with $^{12}$C, $^{14}$N and $^{16}$O along with carbon
burning via $\rm ^{12}C + ^{12}C$ only and the disintegration of
$^{20}$Ne.  Other isotopes and reactions are not explicitly followed.
Reaction rates are taken from Caughlan and Fowler
\shortcite{caughlan1988}.  Opacity tables are those calculated by
Iglesias, Rogers and Wilson (1992) and Alexander and Ferguson (1994).
An Eddington approximation (Woolley and Stibbs 1953) is used for the
surface boundary conditions at an optical depth of $\tau = 2/3$.  This
means that low-temperature atmospheres, in which convection extends
out as far as $\tau \approx 0.01$ (Baraffe et al. 1995), are not
modelled perfectly.  However the effect of this approximation on
observable quantities is not significant in this work (see for example
Kroupa and Tout 1997). We included convective overshooting in this
run of models as described in Pols et al. \shortcite{pols97} 
in order to compare consistently with the models of the Geneva group
which incorporate this mixing process. 
There is no mass-loss added to the $9-20$\msol\
tracks as this plays only a very minor role in the evolution of these
stars. Mass loss for the higher mass stars is included as described in
Dray et al. \shortcite{dray03} according to their NL prescription. 

The most interesting luminosity range which is of relevance to this
paper, and the studies of the progenitors of SNe II-P, corresponds to
the $9-20$\msol\ mass range. For these masses the end points of the
evolutionary tracks at the end of core-carbon burning
in our tracks are very close, but not
identical, to those of the Geneva group \cite{sch92,mey94}.  The
Geneva tracks are calculated at 9, 12, 15 and 20\msol, and we have
reproduced these and all integer values in between. Our tracks between
9-15\msol\ end at a luminosity 0.1$-$0.15\,dex higher than
those of the Geneva group, and at 20\msol\ they are virtually
identical. One should note that when convective overshooting is
included, then 2nd dredge-up does not occur.  Smartt et
al. \shortcite{sma02a} discussed the effects of 2nd-dredge in 7-11\msol\
models without convective overshooting, which predicts that 
stars of this mass can achieve significantly higher  
theoretical luminosities. At this
point however this serves as an illustrative example of the
uncertainties in our understanding of the 
convective and mixing processes in massive stars, and the final luminosities
of the 7-11\msol\ stars cannot yet observationally constrain the models. 
This area
requires more careful scrutiny from both the observational and
theoretical side, and we intend to study the latter in future
publications using our stellar evolutionary code. 
Given the fact that the pre-supernova luminosities in the 9-15\msol\ range
are approximately 0.15\,dex higher than the Geneva tracks, in this paper
we  adopt the conservative policy of incorporating the errors
on the luminosity limits in the estimations of the upper mass limits
as described further below.

\section{Discussion of results for SN~2001du}

\subsection{Reddening and metallicity of the SN~2001du region}
\label{red_metal}
Several methods can be used to estimate the reddening towards the
supernova, and hence towards the progenitor object. These have been
discussed previously in Smartt et al. \shortcite{sma01a,sma02a} and in
Leonard et al. \shortcite{leon02a}, and here we try three different
methods to determine the extinction. The method of using the Na\,{\sc
i}D interstellar absorption lines as a tracer of reddening along the
line of sight is also often employed for nearby SNe. However these
lines are not visible in the Sept. 9.35 spectrum. We have measured the
upper limit to their strength of EW$\la 0.4$\AA\ which is already
suggestive of low interstellar absorption in the range $0.06\la E(B-V)
\la 0.20$ as deduced from the empirical relations between the EW of
the  Na\,{\sc i}D  features and line of sight reddening given in Turatto
et al. \shortcite{turatto03}. 

\begin{figure}
\epsfig{file=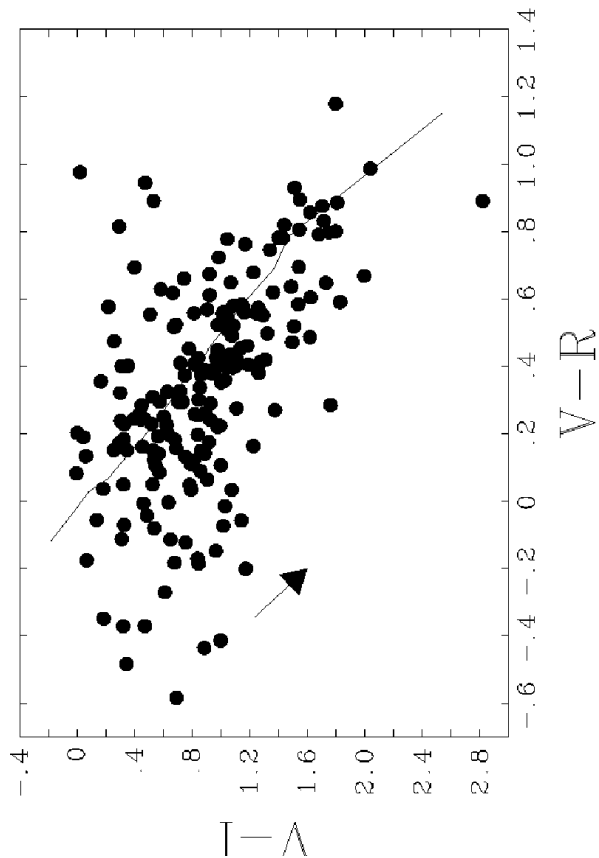,width=8cm}
\caption{Two-colour diagram showing of the supergiants of NGC 1365, in
the vicinity of SN~2001du.  Also shown is the sequence of theoretical
supergiant colours (Bessell 1990) reddened by $E(V-I)=0.155\pm0.028$.
The reddening vector is indicated by the arrow.}
\label{twocolfig}
\end{figure}

\subsubsection{Comparison of the continuum slope with SN~1999em and a black-body function}
\label{discuss_spec}
As discussed in Leonard et al. \shortcite{leon02a} and 
Eastman et al. \shortcite{east96}, a hard upper limit
to the extinction toward a SN II-P in the photospheric stage
can be determined. An arbitrarily, and unphysically, hot 
black-body temperature is compared with the dereddened continuum
of the supernova. If SN~2001du is dereddened by $E(B-V)>$0.6
then even a function with $T_{\rm BB}=10^9$\,K is unable to 
reproduce the slope of the continuum (see Fig.\,\ref{2001du_1999emspec}). 
As pointed out by Leonard et al. 
\shortcite{leon02a} it is desirable to have a spectrum 
taken as early as possible after explosion, because as the 
photosphere cools the upper limit that can be derived will become
less restrictive. Using this method they derived a limit of 
$E(B-V)<0.45$ for SN~1999gi with a spectrum taken 0.62 days 
after discovery. As we discuss below, our earliest 
spectrum of SN~2001du was taken approximately 8 days after discovery. 
This establishes a very secure 
upper limit to the extinction, however in all likelihood the 
actual reddening will be lower, and a more applicable 
$T_{\rm BB}$ will be of the order 10000$-$20000\,K. 
In Fig.\,\ref{2001du_1999emspec} we plot the black-body fluxes
along with suitably dereddened observed continua, and conclude
that the reddening towards SN~2001du is likely to be 
of the order 0.0$-$0.22, with a hard upper limit of $E(B-V)$=0.6. 

\begin{table*}
\begin{minipage}{126mm}
\caption{Limits on the bolometric magnitude and luminosity of the progenitor
star of SN~2001du using the 5$\sigma$ limits from the $V_{555}$ and
$I_{814}$ images.}
\label{bcorrtab}
\begin{tabular}{lcrrrrrrr}
\hline\hline
 & & & $M_{V}=-7.5$ & & & & $M_{I}=-8.2$ \\
\cline{3-5}\cline{7-9}\\
Spectral & $\mathrm{T_{eff}}$ & BC & $\mathrm{M_{bol}}$ & log$L/L_{\sun}$& & BC+$(V-I)_{0}$ & $\mathrm{M_{bol}}$& log$L/L_{\sun}$ \\
Type & (K)\\
\hline
O9 & 32000 & $-$3.18  & $-$10.68 & 6.17 & & $-$3.65  & $-$11.85 & 6.64 \\
B2 & 17600 & $-$1.58  &  $-$9.08 & 5.53 & & $-$1.78  & $-$9.98  & 5.89 \\
B5 & 13600 & $-$0.95  &  $-$8.45 & 5.28 & & $-$1.00  & $-$9.20  & 5.58 \\
B8 & 11100 & $-$0.66  &  $-$8.16 & 5.16 & & $-$0.64  & $-$8.84  & 5.43 \\
A0 &  9980 & $-$0.41  &  $-$7.91 & 5.06 & & $-$0.33  & $-$8.53  & 5.30 \\
A2 &  9380 & $-$0.28  &  $-$7.78 & 5.01 & & $-$0.14  & $-$8.34  & 5.23 \\
A5 &  8610 & $-$0.13  &  $-$7.63 & 4.95 & &  0.12    & $-$8.08  & 5.13 \\
F0 &  7460 & $-$0.01  &  $-$7.51 & 4.90 & &  0.40    & $-$7.80  & 5.02 \\
F2 &  7030 &    0.00  &  $-$7.50 & 4.90 & &  0.47    & $-$7.73  & 4.99 \\
F5 &  6370 & $-$0.03  &  $-$7.53 & 4.91 & &  0.55    & $-$7.65  & 4.96 \\
F8 &  5750 & $-$0.09  &  $-$7.59 & 4.93 & &  0.63    & $-$7.57  & 4.92 \\
G0 &  5370 & $-$0.15  &  $-$7.65 & 4.96 & &  0.69    & $-$7.51  & 4.90 \\
G2 &  5190 & $-$0.21  &  $-$7.71 & 4.98 & &  0.77    & $-$7.43  & 4.87 \\
G5 &  4930 & $-$0.33  &  $-$7.83 & 5.03 & &  0.78    & $-$7.42  & 4.86 \\
G8 &  4700 & $-$0.42  &  $-$7.92 & 5.06 & &  0.73    & $-$7.47  & 4.88 \\
K0 &  4550 & $-$0.50  &  $-$8.00 & 5.10 & &  0.74    & $-$7.48  & 4.89 \\
K2 &  4310 & $-$0.61  &  $-$8.11 & 5.14 & &  0.79    & $-$7.41  & 4.86 \\
K5 &  3990 & $-$1.01  &  $-$8.51 & 5.30 & &  1.09    & $-$7.11  & 4.74 \\
M0 &  3620 & $-$1.29  &  $-$8.79 & 5.41 & &  0.88    & $-$7.32  & 4.82 \\
M2 &  3370 & $-$1.62  &  $-$9.12 & 5.54 & &  0.82    & $-$7.38  & 4.85 \\
M5 &  2880 & $-$3.47  & $-$10.97 & 6.28 & &  0.67    & $-$7.53  & 4.91 \\
\hline
\end{tabular}
\end{minipage}
\end{table*}

The supernova SN~1999em is a very well sampled nearby type II-P event
and has been studied in detail by Baron et al. \shortcite{bar00},
Hamuy et al. \shortcite{ham01}, Leonard et al. \shortcite{leon02b} and
Elmhamdi et al. \shortcite{elm02}.  The latter three studies have
extensive optical spectroscopy during the plateau phase.  Hamuy et
al. have kindly made their data available in electronic format on the
{\sc suspect}\footnote{http://tor.nhn.ou.edu/$\sim$suspect/} archive
website, and we have access to the 
Elmhamdi et al. electronic data. The slopes of the continua 
in the Elmhamdi et al. and the Hamuy et al. spectra 
around day +8 after explosion are very similar, and 
in Fig.\,\ref{2001du_1999emspec} the similarity with SN~2001du 
is illustrated. 
The time of explosion of SN~1999em is taken as
$t_{0}=2451478.8$ JD from the EPM analysis \cite{ham01}, hence the
1999 November 14 spectrum corresponds to day +18.
Fig.~\ref{2001du_1999emspec} shows the excellent match between the
+18 day spectrum of SN~1999em and our SN~2001du ESO spectrum taken on
2001 September 9.35. We attempted to match the SN~2001du spectra with
the Hamuy et al. data for +13 (November 9) and +23 (November 19) days
and applied variable reddening to the spectra of SN~2001du but could
not consistently match the continuum slope, the strength of the
absorption lines and the H$\alpha$ P-Cygni feature. In addition we
varied the absolute linear scaling factor applied to the SN~2001du
spectrum (to account for distance and intrinsic magnitude differences)
and the amount of dereddening (assuming R=3.1) when matching it to the
+18 day SN~1999em spectrum. We found that any extra reddening added to
SN~2001du was incompatible with a good match and even applying an
additional reddening of $E(B-V)=0.1$ gave poor results. Hence
the spectrum of SN~2001du taken on 2001 September 9.35 UT is virtually
identical with that of SN~1999em at +18 days after explosion and
implies an identical reddening.  Baron et al. \shortcite{bar00} have
compared the early time optical and UV spectra of SN~1999em with a
non-LTE model atmosphere fit. They simultaneously determine a
temperature and a low value of reddening of $E(B-V)=0.10\pm0.05$, and
argue that increasing the value of $E(B-V)$ will not lead to a
consistent fit to all the features in the spectra. In summary, the
continua and line strengths of SN~1999em and SN~2001du on +18 days are
virtually identical, and any further artificial reddening of SN~2001du
leads to significant inconsistencies. Hence our estimate of reddening
towards SN~2001du using this method is $E(B-V)=0.1\pm0.05$.

\subsubsection{Estimates of reddening from nearby H\,{\sc ii} regions}
The abundance gradient in NGC1365 has been studied by 
Roy \& Walsh \shortcite{roy97}
who observed the nebular lines of 55 H\,{\sc ii} regions. For 
each H\,{\sc ii} region they determined an extinction at H$\beta$, 
which is the combined value of Galactic reddening ($A_V=0.16$), 
and extragalactic extinction. Their nebular region RW21 is 
virtually coincident with the star formation region around
the position of the supernova, and the centre of this H\,{\sc ii}
region is estimated to be approximately 4$''$ from SN~2001du.
Hence it is likely that this nebulosity is directly 
associated with the ionizing hot stars in the vicinity of SN~2001du
i.e. the bright clustered sources in the $U_{336}$ image in 
Fig.\,1(a). Roy \& Walsh \shortcite{roy97} measured 
a logarithmic extinction at 4861\AA\ of 
$c({\rm H}\beta)=0.83\pm0.06$ for RW21 and using an
extinction law of $A(4861)/A(V)=1.16$  \cite{card89} 
this translates into $A_V=1.8\pm0.1$, or $E(B-V)=0.58\pm0.03$ 
(assuming R=3.1). There are two other H\,{\sc ii} regions lying 
in the same spiral arm at the end of the central bar in NGC1365
both at distances of approximately $15''$ from the supernova
(RW19 and RW20). These have a virtually identical extinction of 
$E(B-V)=0.53\pm0.03$, hence the extinction toward the
ionizing regions near the SN position which produce the nebular lines 
is of order $E(B-V)=0.58\pm0.03$. However as discssed by Calzetti et al.
\shortcite{cal94} and Calzetti \shortcite{cal97} the extinction measured
in starforming regions from the nebular lines is generally a factor of 
two higher than that of the underlying stellar continuum. Hence 
the H\,{\sc ii} region reddening is not likely to be directly applicable
to the SN~2001du reddening as the supernova it is not coincident with the 
position of the UV-bright ionizing O-stars. We hence assume that a
reddening a factor of two lower should be applicable to the stars
near the RW21 region i.e. $E(B-V)=0.29\pm0.03$. While this 
may appear somewhat arbitrary, we should get a reliable mean
when combining it with the other extinction estimates . 

Roy \& Walsh determined the oxygen abundance in their H\,{\sc ii}
regions from an analytic calibration of the $R_{23}$ ratio 
$R_{23}$=([OII]$\lambda$3727+[OIII]$\lambda$$\lambda$4959,5007)/H$\beta$.
While this method usually leads to a reasonably consistent estimation of 
how abundances differentially vary within a galaxy, the absolute 
abundance determined for any particular $R_{23}$ ratio is highly 
dependent on the calibration employed. Recently 
Smartt et al. \shortcite{sma01b} and Trundle et al. 
\shortcite{tru02} have discussed various literature 
calibrations of the $R_{23}$ ratio, and compared 
extragalactic nebular abundances
directly with massive stars born within the H\,{\sc ii} regions. 
They suggest that the calibrations of Pilyugin \shortcite{pil02} and 
McGaugh \shortcite{mcg91} (the latter parameterized by Kobulnicky et al. 
\shortcite{kob99}) give the best agreement for oxygen abundances in 
the gas and in massive B-type supergiants. Applying these calibrations
to the $R_{23}$ ratio of RW21 results in an oxygen abundance 
$12 + \log{\rm (O/H)}$=8.5\,dex (from the Pilyugin calibration), 
and 8.8\,dex (from McGaugh). Hence within the uncertainty of the
absolute calibration of this method, the H\,{\sc ii}
region RW21 is approximately solar in metallicity; assuming 
the solar oxygen abundance is 8.83\,dex \cite{grev98}. 

\subsubsection{The reddening towards surrounding stars in NGC1365} 
\label{discuss_phot}
In this case the reddening can be measured from 3-colour photometry of
the bright supergiants in the area around the supernova, and one then
assumes that this is representative of the reddening toward the
supernova itself. Photometry in three filters (F555, F675W and F814W
transformed to standard $VRI$) was measured for 190
stars  on the PC chip in the vicinity of SN~2001du (to a radius of
23\arcsec from the SN position). The locus of the observed stars, in
a two-colour plane ($V-I, V-R$) was compared with a theoretical
supergiant colour sequence from Bessell \shortcite{bes90} for galactic
supergiants.  The reddening was determined by ascertaining the
displacement between the observed stars and the theoretical sequence in
the two-colour plane.  A weighting factor was included in the
calculation, in favour of those stars with more accurately determined
magnitudes. The reddening was determined to be $E(V-I)=0.16\pm0.03$.
Fig.\,\ref{twocolfig} shows the two-colour diagram of the stars
imaged in the vicinity of SN~2001du; also shown is the theoretical
supergiant sequence appropriately reddened.  Utilising a reddening law
of $R_{V}=A_{V}/E(V-I)=2.45$ \cite{silb99} yielded a value of
$A_{V}=0.38$.  In determining a reddening towards the 
Cepheid population of NGC1365 (in order for distance determinations), 
Silbermann et al. \shortcite{silb99} measured a value of 
$A_{V}=0.40$. Within the errors this is identical to our 
measurement in the field of SN~2001du, and suggests that the reddening
towards this part of the galaxy is not significantly different to that
derived by Silbermann et al., and is not prohibitively high. 
A standard reddening law of
$R_{V}=A_{V}/E(B-V)=3.1$ would imply a value of $E(B-V)=0.12\pm0.02$
is appropriate for the line-of-sight to SN~2001du from this method.

\subsection{A luminosity and mass limit for the progenitor 
of SN~2001du}
\label{discuss_limits}

\begin{figure}
\epsfig{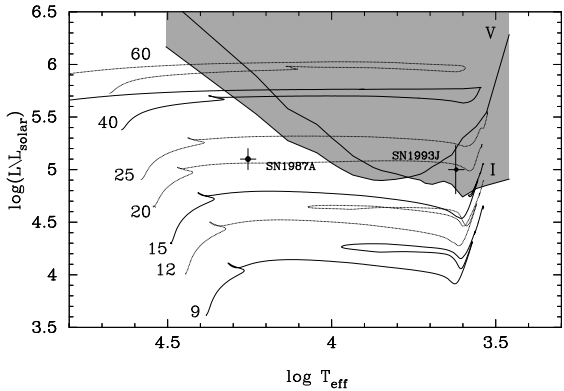}
\caption{SN~2001du: Luminosity limits from pre-explosion images in the $V$ and
$I$ are plotted as thick solid lines.  The shaded region is where a
progenitor would have been detected in at least one filter. Overlaid
are the evolutionary tracks as described in Sect. 3 
for stars with main sequence masses
$\mathrm{9-60M_{\sun}}$, for solar metallicity.  
The locations of the progenitors of SN~1987A and SN~1993J are also shown.}
\label{01duhrdfig}
\end{figure}

\begin{figure}
\epsfig{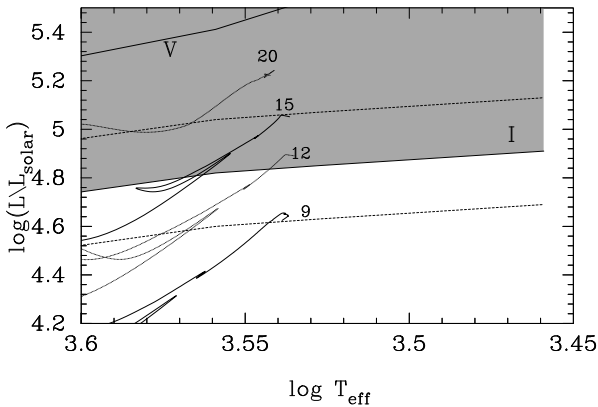}
\caption{SN~2001du: End stages of the evolutionary tracks for 
stars with initial masses in the range 9$-$20\msol.  The
limiting luminosities are shown as the thick solid lines.  Uncertainty
limits, on the I-band limiting luminosity are shown as the dashed
lines.  The limiting luminosity including the expected error
constrains a robust upper mass limit of 15\msol}
\label{01duhrdclose}
\end{figure}

The three methods above give reassuringly consistent measurements for
the reddening towards SN~2001du, and presumably its progenitor
star. Hence we adopt a mean value of these estimates of
$E(B-V)=0.17\pm0.09$ and (assuming $R_{V}=3.1$)
$A_{V}=0.53\pm0.26$. Smartt et al. \shortcite{sma01a,sma02a} discuss
the general technique employed to determine an upper luminosity limit
for the progenitor for it not to have been detected on pre-explosion
images.  The pre-explosion $U_{336}$ archive image was not
sufficiently deep to place any significantly constraining upper
luminosity limits for the progenitor star and is not considered
further in the following discussion.  The $V_{555}$ and $I_{814}$
5$\sigma$
detection limit magnitudes were transformed to $V=24.4$ and $I=23.5$
using the transformations of Holtzman et al. \shortcite{holtz95b}.
The corresponding absolute magnitudes ($M_{V}$ and $M_{I}$) were
calculated by correcting for the extinction, determined earlier, and
adopting a distance modulus to NGC 1365 of $(m-M)=31.39\pm0.26$
\cite{ferr00}.  The limiting absolute magnitudes were determined to
be: $M_{V}=-7.5$ and $M_{I}=-8.2$.  Bolometric magnitudes, for each
waveband, were determined by applying bolometric corrections, and an
appropriate colour correction $(V-I)_{0}$ for $M_{I}$, both taken from
Drilling \& Landolt \shortcite{drlan00}.  The luminosity limits, as a
function of temperature, were calculated from the bolometric
magnitudes assuming a solar bolometric magnitude of $M_{\rm
bol}=+4.74$, and are listed in Table\,\ref{bcorrtab}. These luminosity
limits were overlaid on an HR diagram with our evolutionary tracks (as
described in Section \ref{stellar_models}).  An HR diagram showing the
exclusion region defined by the luminosity upper limits is shown in
Fig.\,\ref{01duhrdfig}.  A closer view of the end-points of the tracks
in the red supergiant region, and the exclusion limit defined by the
$I$ band sensitivity limit is further shown in
Fig.\,\ref{01duhrdclose}.  The expected magnitude errors in distance
modulus ($\pm$0.26), reddening ($\pm$0.26), sensitivity limits
($\pm$0.2), and BC and intrinsic stellar colours ($\pm$0.4) are
combined in quadrature giving an uncertainty on $M_{\rm bol}$ of
$\pm$0.58 mag. As we pointed out in Section\,\ref{stellar_models}, 
the end-points of our tracks in the red-supergiant region 
tend to be systematically higher than the Geneva tracks
by 0.1$-$0.15\,dex. As the observational uncertainties are 
significant we adopt the conservative approach of adding the 
total error to the luminosity upper limit to derive a robust upper
mass-limit. In this case we can say that the upper limit to the
initial mass of the progenitor star of SN~2001du is 15\msol. 

As discussed in Sect.\,\ref{data} we believe it unlikely that star
B is a single point source which is precisely coincident with the 
the supernova and is an unlikely progenitor candidate. 
However if it 
was the progenitor it would have M$_{\rm V}=-6.8$, and 
$V-I \leq 1.4$. This would constrain it to be just below the 
exclusion region in Fig.\,\ref{01duhrdfig} and to be bluer than 
approximately K2 spectral type i.e. cooler than 
$\log T_{\rm eff} \lesssim 3.7$, and with a mass in the region
15-20\msol. Hence it would not have been an
M-type supergiant, but an unperturbed early K-type could provide a large
enough hydrogen envelope to sustain a plateau phase of $\sim$100 days
(see Table\,\ref{proglist_all}), and is thus not inconsistent with 
the supernova evolution. Van Dyk et al. \shortcite{vandyk03b} 
have also reanalysed the site of SN~2001du using our images of the 
supernova, and come to the same conclusion that Star B is not the 
progenitor. The upper mass limits derived, again repeating our method, 
are in agreement with those derived here (13M$\sol^{+7}_{-4}$), 
although we have reduced the uncertainties using our finer
grid of stellar evolutionary tracks. 

\section{Discussion of results for two other II-P supernovae}

\begin{figure*}
\epsfig{file=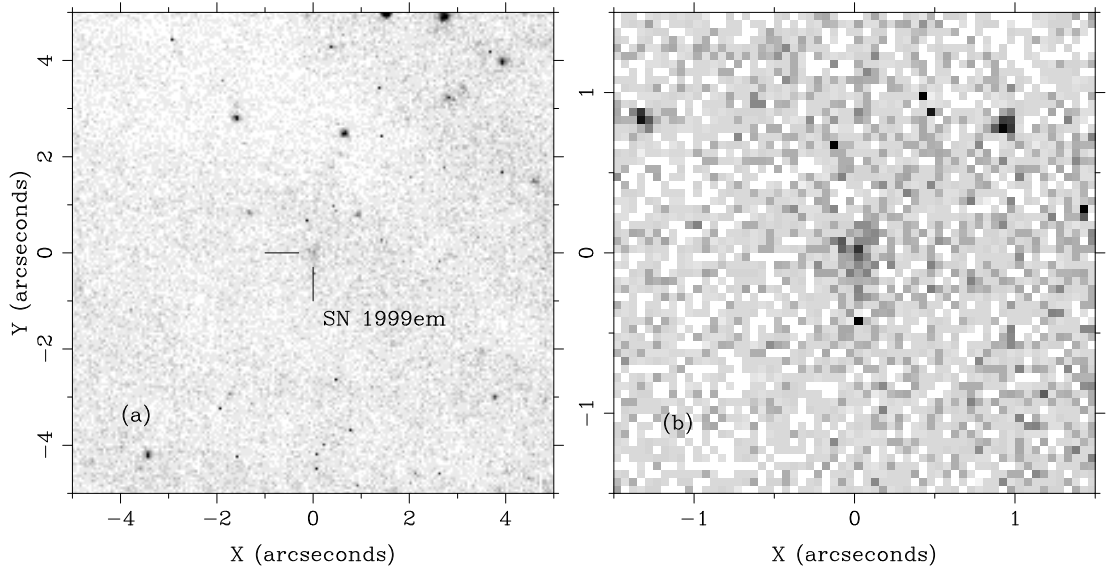,width=16cm}
\caption{{\bf (a):}
A WFPC2 F555W image of the SN\,1999em taken on 2002 January 15,
some +810 days after explosion. The position of the supernova was identified
from the high-resolution WHT images presented by Smartt et al. 
(2002a). {\bf (b):} Close up of the 
faint source at the position of SN~1999em, which is not a simple 
point source (compare with the other two point sources at 
$y\simeq0.6$). It appears as two peaks superimposed on a faint
diffuse background. 
However the region around the SN position is uncrowded and relatively 
clear of contaminating bright sources 
within the $0.7''$ seeing disk of the pre-explosion frames of 
Smartt et al. (2002a). This validates the use of these lower resolution
frames in setting a luminosity and mass limit for the progenitor star. 
}
\label{1999emwfpc2}
\end{figure*}

\subsection{SN~1999em}

\begin{figure}
\epsfig{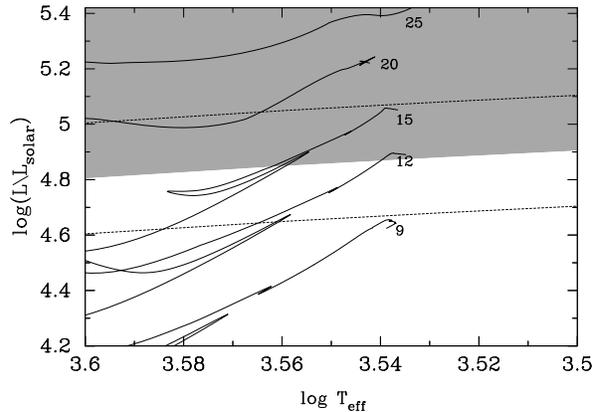}
\caption{SN~1999em: updated exclusion region diagram for the 
progenitor of SN~1999em (from combined $R$ and $I$ limits), 
using the new distance estimate for 
NGC1637 of 11\,Mpc. Our evolutionary tracks as discussed in 
Sect.\,3  are plotted along with the uncertainties on the
luminosity limits. The new distance increases the upper mass 
limit to 15\msol.}
\label{sn1999emhrd}
\end{figure}

In Smartt et al. \shortcite{sma02a} we discussed the prediscovery
images of SN~1999em which were taken from the ground with the
Canada-France-Hawaii telescope at a resolution of $0.7''$.  As with
SN~2001du and SN~1999gi (see Section\,\ref{discuss_1999gi}) there was no
detection of a progenitor star at the position of SN~1999em, and the
sensitivity limits of the $VRI$ frames allowed an upper mass limit to
be derived. Similar methods as discussed above led us to determine an
upper mass limit of approximately 12\msol. 
However these conclusions are based
on the assumption that the ground-based images do not have 
contaminating objects within the seeing PSF of $0.7''$. As
shown in the case of SN~1999gi, which exploded in a fairly
large OB-association, it is quite possible that ground-based
resolution images would not be adequate to resolve the 
individual massive stars. For that reason we have reobserved the
environment of SN~1999em with WFPC2 during our Cycle\,10 
program GO9041. The main goal of this observation was to 
investigate the immediate environment surrounding 
the progenitor of SN~1999em, and determine 
if a contaminating star cluster 
would compromise the result. SN~1999em was observed with 
WFPC2 on 2002 January 15 through the filters F555W, F675W
and F814W. The F555W image is shown in Fig.\,\ref{1999emwfpc2}, 
and we have identified the position of the supernova through
matching the brightest single stars in the field of the WFPC2
data and the WHT image of Smartt et al. \shortcite{sma02a}. 
A geometric transformation between the two frames results in 
a residual uncertainty of approximately $0.1''$ in the SN position.
The WFPC2 image shows a faint object at the position of
the supernova, which is not a clean point source. Two 
peaks appear, superimposed on a quite faint extended region
of flux. Although this diffuse emission is faint it is clearly 
visible in both the F555W, and the F675W but not the F814W filter. 
We estimate the position of the brighter peak at the centre
of Fig\,7a is within $0.08''$ of the supernova centroid, which 
is within the errors of our differential astrometry. The extended weak 
flux is approximately $0.5''$ in diameter, or 25\,pc (assuming 
the distance discussed below). This is much too far to be due to a
light echo from SN~1999em and is likely a small star cluster
which hosted the progenitor star. The magnitude of this object as a whole
(from aperture photometry using a radius of $0.3''$) is
$V_{555}=23.7\pm0.2$, which is a combination of the residual 
supernova magnitude and the underlying association. This would imply 
that the host cluster has $M_{V} > -6.1$, which is below the 
detection limit of the pre-explosion frames ($M_{V}=-6.49$; Smartt et
al. 2002a). Hence the size and luminosity of the cluster is 
fairly typical for the smallest young associations
found in spirals and in the Local Group
e.g. Bresolin et al. \shortcite{bres96,bres98}. 
For this paper, the most important conclusion is that the images
confirm that within the seeing disk of our
prediscovery CFHT images there are no other {\em bright} contaminating 
objects.

Leonard et al. (2003, in preparation) have observed the galaxy
NGC1637 on several epochs with HST/WFPC2 during Cycle\,10 to measure a
Cepheid distance to the galaxy. This is the first attempt to compare a
Cepheid distance to the expanding photosphere method as applied to the
type\,II supernovae, and the Cepheid distance measurement is
$D\sim$11\,Mpc (D. Leonard
and N. Tanvir, private communication). Hence we have revised the 
exclusion diagram presented in Smartt et al. (2002a) with this 
new distance. As the HST/WFPC2 images confirm that no contaminating 
stellar sources exist within the ground-based seeing disk that 
would compromise the point source sensitivity limits, the 
diagram shown in Fig.\,\ref{sn1999emhrd} should now be a robust result. 
Applying the error estimates and (as in Smartt et al. 2002a) 
the same conservative criteria as applied to SN~2001du of using
the highest mass that falls within the upper bounds of the errors, 
we now derive an upper mass limit of 15\msol. 
The fact that the underlying association or cluster,
in which the progenitor was born, was small adds further weight to 
our argument that the progenitor could not have been very massive. 
Assuming a typical initial mass function, 
it is statistically unlikely that a cluster of this size would have
given birth to very massive stars of greater than 20\msol. 
The new limit derived is tabulated in Table\,\ref{proglist_all}
for comparison with other known progenitors. 

\begin{figure}
\epsfig{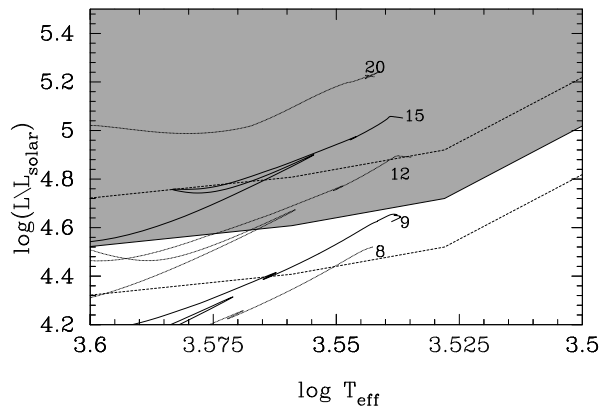}
\caption{SN~1999gi: End points of stars with initial 
masses in the range 8$-$20\msol\ using our evolutionary 
tracks as described in Sect. 3. The
limiting luminosities (from the F606W filter)
are shown as the thick solid lines and 
the uncertainties on these are the dashed lines.  This 
gives a robust upper limit of 12\msol. }
\label{1999giclose}
\end{figure}

The unexpected faint magnitude of the supernova has implications 
for the Elmhamdi et al. \shortcite{elm02} conclusions. Even assuming
the total magnitude measured (i.e. supernovae plus underlying cluster)
to be an upper limit to the SN~1999em flux itself, 
it is  significantly fainter than
the last reported detection by Elmhamdi et al. \shortcite{elm02}
of $V=20.05\pm0.18$ 
on 2001 March 16. Elmhamdi et al. have suggested that their
last photometric point is fainter than the expected
linear decay (extrapolated from previous points) 
of $\gamma_v\sim0.97$ mag\,(100d)$^{-1}$ by
approximately 30\%,  and have interpreted this as a sign 
of dust formation. If we extrapolate the linear decay 
of Elmhamdi et al. (from the last point on 2001 February 01), 
then the SN~1999em should be V$\simeq$23.0 (at day +810) in the HST
WFPC2 images. As it is 0.7$^{m}$ fainter (assuming the 
F555W bandpass is close to $V$) this is supporting
evidence that the deviation they reported is real, and continues
even more severely at late times. Elmhamdi et al. also
suggested that the [O\,{\sc i}] line profile variations after day $\sim$465,
were indicative of dust formation in the cooling supernova. 
This faint WFPC2 magnitude of the SN~1999em would seem to add 
further support to that. The multi-epoch 
images of Leonard et al. (G0~9155) to monitor the Cepheid 
variable population also contain the SN~1999em on the field of 
view, hence this will allow them to probe the late time 
lightcurve in F555W and F814W, and dust formation scenario, 
in much more detail.

\begin{figure}
\epsfig{file=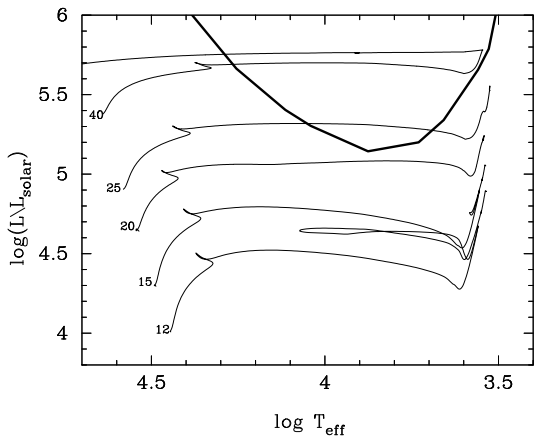,width=8cm}
\caption{Estimate of the lower mass limit for the progenitor of 
SN~1997bs. The object was detected at $M_{V}\simeq-8.1$
(in F606W; Van Dyk et al. 2000), and the bolometric luminosity locus as
a function of $T_{\rm eff}$ (from the supergiant table in 
Drilling and Landolt 2000) is plotted. 
Overlaid are evolutionary tracks for stars with main sequence masses
$\mathrm{12-40M_{\sun}}$, for solar metallicity. The progenitor must have
been somewhere on this line prior to explosion, hence an approximate 
lower limit to the mass is M$>20$\msol.}
\label{sn1997bsfig}
\end{figure}

\subsection{SN~1999gi}
\label{discuss_1999gi}
In Smartt et al. \shortcite{sma01a} we presented the HST archive
pre-explosion images for SN~1999gi along with images of the supernova
taken 14 months later in order to astrometrically 
determine the supernova position 
precisely. There was no detection of a progenitor
star in either the F300W or F606W filter, and using the methods
described above in Section\,\ref{discuss_limits} we derived an upper
mass limit for the progenitor star of 9$^{+3}_{-2}$\msol, which
assumed it was a red supergiant in the final stages of its
evolution. However in this paper we assumed a distance of
$7.9\pm2$\,Mpc, and recently Leonard et al. \shortcite{leon02a} have
determined an improved distance to SN~1999gi using the expanding
photosphere method. They estimate a distance of $11.1\pm2$\,Mpc,
and have also carried out a careful comparison of reddening towards the 
supernova from four different methods (which we followed in 
Sect.\,\ref{red_metal}). Leonard et al. estimate a slightly higher value of 
$E(B-V)=0.21$ compared to the $E(B-V)=0.15$ used in 
Smartt et al. (2001a), and derive an upper mass limit for the star of 
15$^{+5}_{-3}$\msol\ on this basis. 

We have recalculated the luminosity and mass limits using the new
distance of Leonard et al., and the higher reddening, but using the
better sampled tracks at our disposal we can reduce the uncertainty in
the mass range of Leonard et al.'s result.
Leonard et al. have estimated the mass limit by adding the 
upper error uncertainty and simply taking the nearest modeled
progenitor mass to that, with an upper limit of the next  highest
mass model in the Geneva tracks. The exclusion region of the
HRD is plotted in Fig.\,\ref{1999giclose}, and applying the
same criteria as for SNe 2001du and 1999em we get an upper
mass limit of 12\msol, which is below our best estimate
{\em and} the uncertainty. The difference between this
limit and the 15$^{+5}_{-3}$\msol\ estimate by Leonard et al. 
is that our tracks end at slightly higher luminosities than
the Geneva models, and also we can rule out the rather large 
range by employing the integer sampled mass tracks. It is 
understandable that Leonard et al. employed a conservative
approach in estimating the upper limit and errors. However
we can certainly rule out the progenitor being a
20\msol\ red supergiant, as this lies significantly above the
detection limits -- it would have a luminosity of $\log
L/L_{\odot}\simeq5.3$, and hence a magnitude of $V_{\rm
606}\simeq23.9$, which would have been a clear (10$\sigma$) detection
on the pre-explosion images. In addition there may be a slight 
discrepancy between our temperatures of pre-supernova M-supergiants
and those in  Fig.\,10 of Leonard et al., where the
effective temperatures of the evolutionary track endpoints 
appear too cool by $\sim$0.05\,dex. 
They use the Geneva tracks (as employed in Smartt et al. 2001a), 
but the plot does not exactly match the tabulated
values i.e. their end-points would be close to $\log {\rm
T_{eff}}$=3.5 in our Fig\,\ref{1999giclose}. If the tracks 
are shifted to the tabulated temperature then their luminosity 
limit of the F606W filter would no longer skim the 20\msol\ 
track, and the 15\msol\ track would be inside the exclusion region. 

We accept the distance and
reddening determinations of Leonard et al.  as more appropriate to
those used in Smartt et al. (2001a), and that Leonard et al.
are being understandably cautious in using the estimate
plus uncertainty to derive a very hard upper limit. But 
the simplest interpretation of Fig.\,\ref{1999giclose}, 
where we use better sampled end points in the tracks, indicates an
upper mass-limit of 12\msol. Clearly these
assumptions rely on how accurate the theoretical stellar evolutionary
models are in predicting the luminosity and temperatures of M-type
supergiants.

\subsection{Progenitors of core-collapse supernovae}
\label{discuss_all}

\begin{table*}
\caption{A list of known supernova progenitors and constraints on their
masses and spectral types 
for supernovae which are certain to be real core-collapse events and 
have been reliably typed. The estimated initial mass of the progenitor 
star is listed as M$_{i}$ and is from direct measurements of the progenitor
star or limits thereon. Compilation values for mass and 
metallicity are taken from Smartt et al. 
(2002a), and updated with results from this paper and 
Smartt et al. (2002b; for SN~2002ap). We estimated the metallicity
of the progenitor of 1997bs simply from its distance from the centre 
of NGC3627, and applied a typical abundance gradient for an Sb galaxy 
from Vila-Costas \& Edmunds (1992). 
The mass of the ejecta in the supernova is M$_{ej}$
and the radii of the progenitors are 
estimated from analysis of the supernova spectra and lightcurve, as 
described in Section\,\ref{discuss_all}. The values for SN~1987A and 
SN~1993J, and SN~1980K are taken from Arnett (1996)
and Woosley et al. (1994). }
\begin{tabular}{llllrccclrll}\hline
SN  & Galaxy  & Distance & Type &  Metallicity &  &  & & \multicolumn{4}{c}{Progenitor properties} \\
                &   &   (Mpc)   &     & (Z/\zsol) &  &  &    & Type &   M$_{i}$/\msol & M$_{ej}$/\msol  & R/\rsol \\
\hline
2002ap & NGC628  & 7.3  & Ic     & 0.5       & ~~ & ~~ & ~~ & WR?         & $<$40 & $2.5-5$ & --  \\
\\                                                          
1997bs & NGC3627 &  11.4 & IIn    & $\sim$1  & ~~ & ~~ & ~~ & ?           & $>$20 & --     & --  \\
1987A  & LMC     &   0.05 & II-pec & 0.5     & ~~ & ~~ & ~~ & B3Ia        & 20    & 15     & 43  \\
1993J  & M81     &   3.6  & IIb    & 2       & ~~ & ~~ & ~~ & G5$-$K0\,Ia & 17   & 3       & $\sim$500  \\
1980K  & NGC6946 &   5.1  & II-L   & 0.5     & ~~ & ~~ & ~~ & ?           & $<$20 & 2      & $\sim$300  \\
\\                                                          
2001du & NGC1365 &   17.9 & II-P   & $\sim$1 & ~~ & ~~ & ~~ & G-M type   & $<$15 & --      & --  \\
1999em & NGC1637 &   11   & II-P   & 1-2     & ~~ & ~~ & ~~ & K-M type   & $<$15 & 5$-$18  & 260$-$1500 \\
1999gi & NGC3184 &   11.1 & II-P   & $\sim$2 & ~~ & ~~ & ~~ & G-M type   & $<$12 & 10$-$30 & 100$-$400 \\

\hline
\label{proglist_all}
\end{tabular}
\end{table*}

In this section we review all of the direct
observational evidence that exists for core-collapse precursors. 
The events for which we have some restrictive information on the
progenitors are listed in Table\,\ref{proglist_all}. Only those
supernovae which are fairly certain to be core-collapse explosions, 
{\em and} have a well defined SNe type attributed are included. 

\begin{figure*}
\epsfig{file=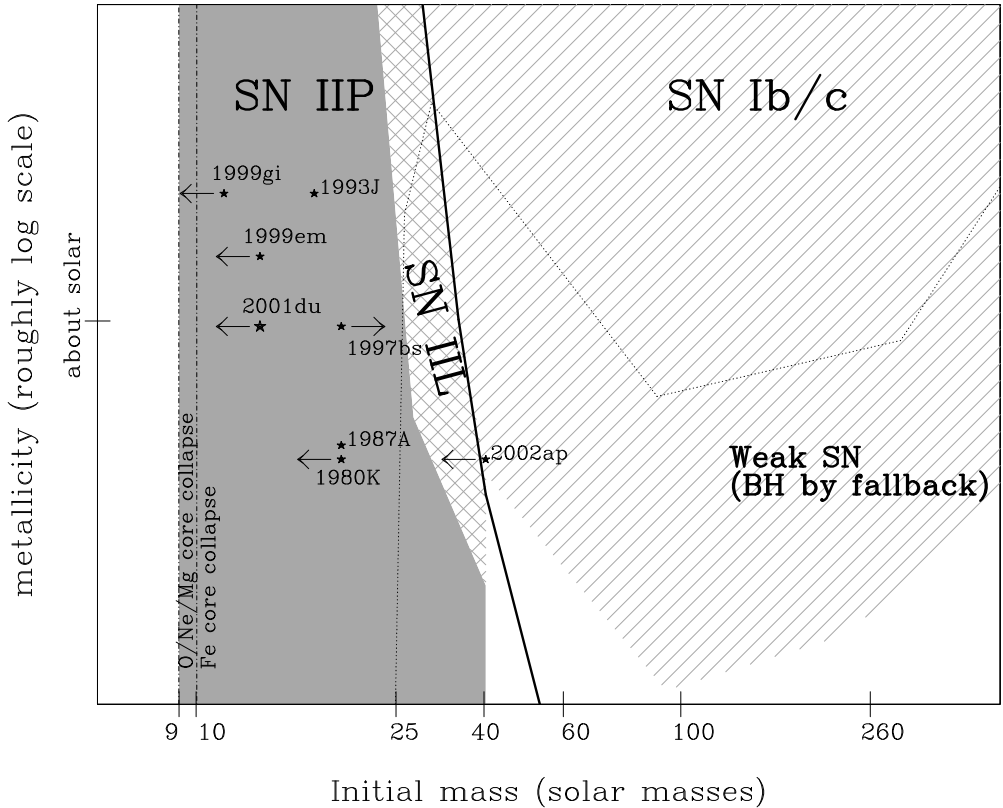,width=16cm}
\caption{The theoretical SNe populations from the models of 
Heger et al. (2003), but ignoring their extremely metal
poor and metal-free objects. All objects below the dotted line should 
be weak SN due to black-hole formation by fall-back of 
most of the material surrounding the iron core. All objects to the
right of the solid black line should have lost their hydrogen
envelope at the point of core-collapse, and hence should be 
hydrogen free SNe. The blank white regions are where no supernova
is seen; below 9\msol\ due to the formation of white dwarfs in
low mass stars; above 40\msol\ due to direct black hole formation.
The observed positions of the supernovae progenitors in Table\,3
are approximately estimated and placed, with arrows indicating 
those with limits. To place the stars in the metallicity
dimension we have assumed a zero-point on the y-axis of 
$\log Z/{\rm Z_{\odot}}$=-1. }
\label{hegercomp}
\end{figure*}

The event SN~1978K  is not included although it had a
progenitor object visible. The supernova itself was 
serendipitously discovered in January 1990 as a peculiar object on an 
archive photographic 
plate \cite{ryder93}. Hence there is only a crude light-curve from archive
plates taken covering that region of sky. Ryder et
al. \shortcite{ryder93} conclude that it was more likely to be a
type\,II event rather than an LBV ``super-outburst'', however this is not
beyond doubt. If it was a supernova we do not know its type, peak
magnitude or very much about the lightcurve. Given the limited
resolution of the photographic surveys (particularly close to the
plate limit) we do not know if the object at the position of SN~1978K
is really a point source, a nebula or cluster
(Ryder et al. quote a distance of 4.5\,Mpc).  These uncertainties
lead us to omit SN~1978K from Table\,\ref{proglist_all} for now, at 
least until some of the questions can be definitively answered.
The event SN~1961V is also not included for similar reasons. 
It is very unclear if it was a genuine supernova 
or an S-Doradus outburst from a very massive star 
like $\eta$~Carinae \cite{hds99}. 
The progenitor object was possibly identified as a star with 
an absolute magnitude $M_{\rm bol}\simeq-11$, which is much too bright
for a stable massive star or $\eta$~Car like object in a quiescent state. 
Hence it was likely to be in outburst during its period as an
$M_{\rm bol}\simeq-11$ object, and given the sparse data available 
we do not have a definitive model for the physics behind the 
eruption. Again because of these uncertainties, we do not discuss this 
further. 

A detection of a progenitor star associated with SN~1997bs in M66
has been claimed by Van Dyk et al. \shortcite{vandyk00}, who originally 
suggested that this event was also not a genuine core-collapse
supernovae but another super-outburst similar to SN~1961V and 
$\eta$~Car, and that the progenitor actually survived the explosion 
in HST images. However Li et al. \shortcite{li02} have observed 
SN~1997bs at a later epoch with HST and observed it to fade considerably, 
and get bluer, compared to the Van Dyk et al. magnitudes. 
With this new information, the conclusion
is that SN~1997bs was indeed likely to have been a subluminous 
IIn supernova, and the
progenitor has not survived. Van Dyk et al. \shortcite{vandyk00}
have estimated the absolute magnitude of the progenitor star to 
be $M_{V} \simeq -8.1$ from the single exposure F606W WFPC2 image. 
As no colour information is available we do not know the temperature
or appropriate bolometric correction to allow the progenitor to be
approximately placed on an HR diagram. However given this
detection in a single band,  a {\em lower} limit can be placed on the mass of 
the progenitor. In Fig\,\ref{sn1997bsfig} we plot the locus of a 
$M_{V}=-8.1$ star as a function of effective temperature and hence
bolometric correction. The progenitor must have existed at some point on
this locus, hence allowing a lower limit to its mass of approximately
$20$\msol. 

The brightest SNe Ic to occur in recent times was SN~2002ap in M74
(=NGC628), and its early lightcurve and spectral evolution was
recently studied by Mazzali et al. \shortcite{mazz02}.  By comparing
explosion models of CO cores combined with spectral synthesis to model
the optical observational data, they estimated an ejecta mass of
between 2.5$-$5\msol. With an assumed (but uncertain within a factor
$\sim$2) remnant mass of roughly 2.5\msol, this would imply the mass
of the CO precursor star should be in the conservative range
2.5$-$10\msol. The final masses of Wolf-Rayet stars of the WC type in
the LMC have been estimated by Crowther et al.  \shortcite{crowth02}
to be in the range 11$-$19\msol, and those in the galaxy in the range
7$-$14\msol. Given the uncertainties, there is reasonable agreement
between these two estimates suggesting a WC type progenitor (of
initial mass $\sim$30\msol) is consistent with the characteristics of
SN~2002ap.  Deep ground-based images of the pre-explosion site of the
supernova were presented by Smartt et al. \shortcite{sma02b}, and no
detection of a massive star was found. This paper ruled out most of
the upper HR-diagram as possible progenitor sites, although this is to
be expected as SNe~Ic show no evidence of the progenitors having
H-rich atmospheres. It is likely that the precursor was a WR star in
the WC phase, and had an initial mass of less than
$\sim$40\msol. However initially lower mass stars which have had their
H-envelopes stripped during mass-transfer in a binary system, rather
than through a radiatively driven stellar wind, still do remain viable
precursors.

The three II-P supernovae discussed above (SNe 2001du, 1999em, 1999gi)
are also listed in Table\,\ref{proglist_all}, along with a similarly
determined upper mass limit for SN~1980K. Pre-explosion images for
this supernova were presented by Thompson \shortcite{thom82}, and we
have recalculated the limit using the same evolutionary tracks as
employed above. The more definite values for SN~1987A and 1993J are
included. The analytical model for the plateau stage of type\,II
supernovae of Popov \shortcite{pop93} produces equations relating the
observed quantities of $t_{\rm p}$ (the plateau duration), $M_{V}$
(absolute visual magnitude at a representative point during the
plateau), and $u_{\rm phot}$ (the expansion velocity of the
photosphere) to three intrinsic properties of the supernova. These are
energy of the explosion and the mass and radius of the initial
envelope of the progenitor star. More detailed hydrodynamic models of
SNe II-P by Litvinova \& Nadyozhin \shortcite{LN85} have produced
similar relations, but with different coefficients. The mass and
initial radius of the ejected stellar envelope for 1999em and 1999gi
has been estimated by Elmhamdi et al \shortcite{elm02}, Hamuy
\shortcite{ham03} and Nadyozhin \shortcite{nad03} using both sets of
equations but differing values for $t_{\rm p}$, $M_{V}$ and $u_{\rm
phot}$.  The mass and radii derived in each paper differ quite
substantially, depending on what values are assumed. The main
differences are the distance assumption, which critically affects
$M_{V}$, and the epoch at which $u_{\rm phot}$ is measured. We have
recalculated the range in feasible values using the most up to date
values for distance and reddening (as discussed above), and get values
of $M_{V}=-16.7$ (1999em) $M_{V}=-15.9$ (1999gi) at approximately
50\,days after discovery (i.e. in the middle of the plateau).  We
assume that $t_{\rm p}\simeq100$\,days and vary $u_{\rm phot}$ between
3000$-$4000\kms. There appears no discrepancy in the actual
measurements of $u_{\rm phot}$ in 1999em between Elmhamdi et
al. \shortcite{elm02}, Hamuy et al. \shortcite{ham01} and Leonard et
al. \shortcite{leon02b}, but there is some difference as to which
epoch is chosen to measure $u_{\rm phot}$ to use in the
calculations. The ranges of ejected mass (M$_{ej}$) and progenitor
radius (R) are tabulated in Table\,\ref{proglist_all}. If we allow
$\sim$2\msol\ to be lost by mass-loss and neutron star formation, then
the mass ranges for the envelope are in reasonable agreement with
our direct estimates for the initial mass of the progenitor stars. The
large range in radius of the envelope is consistent with spectral
types between G0$-$M0.  We do not have enough photometric information
available for SN~2001du to estimate these values for comparison.
However our direct mass limit for SN~2001du is similar to those for
SNe 1999em and 1999gi. Given the very similar spectra of SN~2001du
with both of the latter (see Fig.\,\ref{2001du_1999emspec}), this is
not altogether surprising.  This supports our previous suggestion
\cite{sma02a} that normal SNe II-P result from fairly low mass stars
($<$12\msol) with large radii in the M-supergiant phase that have
undergone only moderate mass-loss. The fact that 1999em and 1999gi
were faint in the x-ray and radio suggests that their progenitor stars
had undergone fairly moderate or low mass-loss. The fluxes in these
wavebands have allowed Schlegel \shortcite{schlegel2001} and Pooley et
al. \shortcite{pool2002} to estimate mass-loss rates for the progenitors of
these two supernova which are the order of
$\dot{M}\sim10^{-6}$\msol\,yr$^{-1}$, with a conservative uncertainty
of a factor of two.  This supports our result that the progenitors
were not very massive stars, as one would expect such stars to have
gone through more drastic mass-loss events and hence to have produced
higher x-ray and radio fluxes. Hence we conclude that moderate mass
stars (of less than approximately 15\msol) which end their lives in
the red supergiant phase tend to give rise to a fairly homogeneous
class of SNe II-P , and initially higher mass stars likely give rise
to more heterogeneous supernovae, although clearly more results are
needed to increase the statistics.

Recent theoretical work by Heger et al. \shortcite{heger03a,heger03b} on the 
the pre-supernova evolution of massive stars as a function
of mass and metallicity, has produced supernova population diagrams. 
These relate the initial mass of a star to the type of supernova 
that is produced, and how that roughly depends on metallicity of the 
original star. The metallicity plays a key role in determining
the mass-loss rate, which in turn is a critical factor in the 
evolution. This has initially been done for single stars, with no 
rotation considered. We have reproduced the figure of 
Heger et al. \shortcite{heger03b}, and ignored the very metal
poor and metal-free regions as they are not applicable to any 
of the local Universe supernovae discussed above. The regions 
between the II-P, II-L and Ib/c regions are delineated by the mass
of the hydrogen envelope which is left just before core-collapse, 
which in turn is dependent on mass-loss history and hence metallicity. 
In Fig.\,\ref{hegercomp}
we have put the positions of the supernovae listed in 
Table\,\ref{proglist_all} to compare with the model predictions. 
We emphasise that at present this is a rather qualitative 
comparison given the approximate metallicity scale proposed on the 
ordinate, and the limits we have for the various nearby supernovae. 
Nevertheless it is certainly valid and essential to compare the 
state-of-the-art observational
and theoretical results in this data starved area. The positions of the
three SNe II-P are in good agreement with the models, as is the 
position of SN~1987A. However SN~1980K was a type II-L but appears in the 
theoretical II-P region. Even
if the metallicity is uncertain by a factor of two, it would be difficult
to envisage pushing this  into the II-L region as it stands. 
This suggests that the progenitor lost more mass than one would 
expect for a 20\msol\ star of metallicity between 0.5$-$1\zsol. 
The extra mass-loss could be accounted for by
interaction within a binary system, with the progenitor losing
its mass through mass-transfer rather than radiatively driven winds. 
SN~1997bs is also likely to have been 
too massive for the II-P region, although given 
that we have only a {\em lower} limit, the actual mass could sit comfortably 
in the II-L area. One serious discrepancy is obvious in that SN~2002ap was
a bright Ic supernova, however the models suggest that these only 
come from either metal-rich stars, or very massive stars with 
$M_{i}>60$\msol. As discussed in Smartt et al. \shortcite{sma02b} 
the initial mass is unlikely to have been as high as $\sim$60\msol, 
hence there is a discrepancy with the position of the weak SN
region. Smartt et al.  \shortcite{sma02b} estimated the metallicity
to be roughly 0.5\zsol, and again even if this was solar the 
position of SN~2002ap would still be discrepant with the 
normal SN Ib/c population region. From the Heger et al. figure 
one would expect that bright Type Ib/c only come from massive
single stars if the metallicity is quite high, and hence they should
be relatively rare. However invoking mass-loss again through
close binary evolution at low metallicities could account for 
objects such as SN~2002ap. The future statistics of masses and 
mass limits for the progenitors of SNe Ib/c will be of crucial 
importance in distinguishing between the two scenarios. 
In summary then it would appear that the theoretical II-P region stretches to 
masses which are a little too high, and that the weak SN region 
should be confined to lower metallicities or much higher masses. 
This is of course a very preliminary comparison, with the 
caveat that only single stellar evolution is included in the 
supernova population diagram. The prospects 
of setting more constraining limits on SNe progenitors or detecting
the stars themselves will allow better observational constraints to 
be placed on these models in the future (see Smartt et al. 2002a 
for a discussion of future prospects). 

\section{Conclusions}

We have investigated HST images of the site of the nearby supernova SN~2001du 
taken 6.6 years before explosion, and compared our results with 
similar pre-explosion investigations of other core-collapse supernovae. 
Our findings are summarised below 

\begin{enumerate}

\item Although there is a 3$\sigma$ detection of a source very 
close to the supernova position in the WFPC2 $V-$band image, it 
is not precisely coincident with the supernova centroid. It is 
slightly beyond the error radius in our analysis and is not 
a clear detection of an unambiguous point source. We conclude
that the progenitor star was likely below the detection limit
of the pre-discovery images  and is not visible on any of the 
frames. Using the 5$\sigma$ sensitivity limits of the images we
estimate that the progenitor likely had an initial mass of $<$15\msol. 

\item SN~2001du was a type II-P, and probably very similar to two
other nearby recent SNe II-P 1999gi and 1999em. We have previously set
similar luminosity and mass limits on the progenitors of those events
\cite{sma01a,sma02a} and in this paper we have revisited those using
new data. A new distance to NGC3184 determined by Leonard et al.
\shortcite{leon02a} slightly changes the mass limit for SN~1999gi to
approximately 12\msol. New HST images of the site of SN~1999em
confirm the validity of our previous results, and using another
new distance increases this upper mass to approximately 15\msol. 

\item Hence all three SNe had very similar mass progenitor stars, 
which is not a surprising result given the photometric and 
spectroscopic similarity of the supernovae themselves. 
The results are consistent (but not uniquely definitive) 
with the idea of SNe II-P
arising in red supergiant progenitors of moderate mass. 

\item We have made first attempts at comparing all the known, direct,
information available on supernova progenitors with theoretical models
of stellar evolution up to the core-collapse stage. These models have
produced supernova population diagrams which relate the initial mass
and metallicity of massive stars to the types of supernovae that they
produce. We find encouraging reasonable agreement between the
observational and theoretical results for the masses of II-P
supernovae progenitors.  However we find discrepancies for a II-L and
bright Ic event.  This is preliminary work that should increase in
statistical significance over the coming years

\end{enumerate}

\section*{Acknowledgments}
This paper was based in part 
on observations made with the NASA/ESA {\em Hubble Space
Telescope}, obtained from the data archive of the Space Telescope
Institute which is operated by the Association of the Universities for
Research in Astronomy Inc., under NASA contract NAS5-26555. These
observations are associated with proposal GO9041.
Some observations were collected at the European Southern
Observatory, Chile, ESO67.D-0594. 
We acknowledge the support given by ASTROVIRTEL, a Project funded by
the European Commission under FP5 Contract No. HPRI-CT-1999-00081. We
thank the ASTROVIRTEL team at ESO/ST-ECF for the new software
developed.  SJS and JRM thank PPARC for financial support in the form of an
Advanced Fellowship award and studentship respectively
and CAT thanks Churchill College for a
fellowship, SB acknowledges support from
the Italian Ministry for Education, University and Research (MIUR)
through grant Cofin MM2001021149-002. 
We thank Peter Meikle for useful discussions and help in
the initial typing of SN~2001du, Bill Januszewski for help in
coordinating and executing our HST observations, and 
Doug Leonard and Nial Tanvir for communicating their new 
Cepheid distance determination to NGC1637 prior to submission of their
paper. Use was made of 
VSNET data for the optical light-curve of SN~2001du, for
which we are very grateful.

\bsp

\label{lastpage}


\begin{thebibliography}{99}
\bibitem[\protect\citename{Aldering et al.\ }{1994}]{alder94}
Aldering G., Humphreys R.M., Richmond M., 1994, AJ, 107, 662
\bibitem[\protect\citename{Alexander \& Ferguson }{1994}]{alexander1994}
Alexander D. R., Ferguson J. W., 1994, ApJ, 437, 879
\bibitem[\protect\citename{Anders \& Grevesse }{1989}]{anders1989}
Anders E., Grevesse, N., 1989, Geochim. Cosmochim. Acta, 53, 197
\bibitem[\protect\citename{Arnett }{1996}]{arnett96}
Arnett D., 1996, Supernovae and Nucleosynthesis, Princeton University 
Press
\bibitem[\protect\citename{Baron et al.\ }{2000}]{bar00}
Baron E., et al, 2000, ApJ 545, 444
\bibitem[\protect\citename{Baraffe et al.\ }{1995}]{baraffe1995}
Baraffe I., Chabrier G., Allard F., Hauschildt P. H., 1995, ApJ, 446, L35
\bibitem[\protect\citename{Bessell\ }{1990}]{bes90}
Bessell M.S., 1990, PASP 102, 1181
\bibitem[\protect\citename{Branch }{2003}]{branch02}
Branch D., 2003, Proc of IAU Symp. No. 212 (2002), eds.
K. A. van der Hucht, A. Herrero, C. Esteban
\bibitem[\protect\citename{Bresolin et al.\ }{1996}]{bres96}
Bresolin F., Kennicutt R.C. Jr., Stetson P.B., 1996, AJ, 112, 1009
\bibitem[\protect\citename{Bresolin et al.\ }{1998}]{bres98}
Bresolin F., et al. 1998, AJ, 116, 119
\bibitem[\protect\citename{Branch et al.\ }{1995}]{branch95}
Branch D., Livio M., Yungelson L. R.; Boffi F. R., Baron E.
1995, PASP, 107, 1019
\bibitem[\protect\citename{Calzetti et al.\ }{1994}]{cal94}
Calzetti D., Kinney A. L., Storchi-Bergmann T, 1994, ApJ, 429, 582
\bibitem[\protect\citename{Calzetti }{1997}]{cal97}
Calzetti D., 1997, AJ, 113, 162
\bibitem[\protect\citename{Cardelli et al.\ }{1989}]{card89}
Cardelli J.A., Clayton G.C., Mathis J.S., 1989, ApJ 345, 245
\bibitem[\protect\citename{Caughlan \& Fowler }{1988}]{caughlan1988}
Caughlan G. R., Fowler W. A., 1988, At. Data Nucl. Data Tables, 40, 284
\bibitem[\protect\citename{Chevalier }{1976}]{chev76}
Chevalier, R. A. 1976, ApJ 207, 872
\bibitem[\protect\citename{Crowther et al.\ }{2002}]{crowth02}
Crowther  P. A., Dessart  L., Hillier  D. J., Abbott  J. B., 
Fullerton, A. W., 2002, A\&A, 392, 653
\bibitem[\protect\citename{Dolphin }{2000a}]{dolph00a}
Dolphin A.E., 2000a, PASP, 112, 1383
\bibitem[\protect\citename{Dolphin }{2000b}]{dolph00b}
Dolphin A.E., 2000b, PASP, 112, 1397
\bibitem[\protect\citename{Dray et al.\ }{2003}]{dray03}
Dray, L.M., Tout, C. A., Karakas, A. I.,Lattanzio, J.C., MNRAS 2003 in press 
\bibitem[\protect\citename{Eastman et al.\ }{1996}]{east96}
Eastman R.G., Schmidt B.P., Kirshner R., 1996, ApJ, 466, 911
\bibitem[\protect\citename{Eggleton }{1971}]{eggleton1971}
Eggleton P. P., 1971, MNRAS, 151, 351
\bibitem[\protect\citename{Eggleton }{1972}]{eggleton1972}
Eggleton P. P., 1972, MNRAS, 156, 361
\bibitem[\protect\citename{Eggleton }{1973}]{eggleton1973}
Eggleton P. P., 1973, MNRAS, 163, 279
\bibitem[\protect\citename{Elmhamdi et al.\ }{2003}]{elm02}
Elmhamdi A., et al., 2003, MNRAS, in press, astro-ph/0209623
\bibitem[\protect\citename{Evans }{2001}]{evans01}
Evans R., 2001, IAU Circ. 7690
\bibitem[\protect\citename{Fassia et al.\ }{2001}]{fass01}
Fassia A., 2001, MNRAS, 325, 907
\bibitem[\protect\citename{Ferrarese et al.\ }{2000}]{ferr00}
Ferrarese L., et al., 2000, ApJSS, 128, 431
\bibitem[\protect\citename{Filippenko\ }{1997}]{fili97}
Filippenko A.V., 1997, ARA\&A, 35, 309
\bibitem[\protect\citename{Grevesse \& Sauval\ }1998]{grev98}
Grevesse N., Sauval A.J., 1998, Space Science Rev., 85, 161
\bibitem[\protect\citename{Hamuy et al.\ }{2001}]{ham01}
Hamuy M., et al, 2001, ApJ, 558, 615
\bibitem[\protect\citename{Hamuy\ }{2003}]{ham03}
Hamuy M., 2003, ApJ, in press, astro-ph/0209174
\bibitem[\protect\citename{Heger et al.\ }{2003a}]{heger03a}
Heger A., Woosley S.E., Fryer S.L., Langer N., 2003a, to appear in
Proc. of the ESO/MPA/MPE Workshop ``From Twilight to Highlight: The
Physics of Supernovae'', eds. W. Hillebrandt, B. Leibundgut  (Springer-Verlag)
in press, (astro-ph/0211062)
\bibitem[\protect\citename{Heger et al.\ }{2003b}]{heger03b}
Heger A.,  Fryer S.L., Woosley S.E., Langer N., Hartmann D.H., 
2003b, ApJ, submitted, astro-ph/0212469
\bibitem[\protect\citename{Humphreys et al.\ }{1999}]{hds99}
Humphreys R.M., Davidson K., Smith N., 1999, PASP, 111, 1124
\bibitem[\protect\citename{Holtzman et al.\ }{1995a}]{holtz95a}
Holtzman J.A., et al., 1995, PASP, 107, 156
\bibitem[\protect\citename{Holtzman et al.\ }{1995b}]{holtz95b}
Holtzman J.A., et al., 1995, PASP, 1065
\bibitem[\protect\citename{Iglesias et al.\ }{1992}]{iglesias1992}
Iglesias C. A., Rogers F.J., Wilson B.G., 1992, ApJ, 397, 717
\bibitem[\protect\citename{Kobulnicky et al.\ }1999]{kob99}
Kobulnicky, H.A., Kennicutt, R.C., Pizagno, J.L., 1999, ApJ 514, 544
\bibitem[\protect\citename{Krist \& Hook\ }{1999}]{kh99}
Krist J., Hook R., 1999, The Tiny Tim Users Guide, wwww.stsci.edu/software/tinytim
\bibitem[\protect\citename{Kroupa \& Tout\ }{1997}]{kroupa1997}
Kroupa P., Tout C. A., 1997, MNRAS, 287, 402
\bibitem[\protect\citename{Drilling and Landolt.\ }{2000}]{drlan00}
Drilling J.S. and Landolt A.U., 2000, in Allen's Astrophysical Quantities, ed. A.N.Cox(4th ed.; New York: AIP)
\bibitem[\protect\citename{Lentz et al.\ }{2001}]{lentz01}
Lentz E.J., et al., 2001, ApJ, 547, 406
\bibitem[\protect\citename{Leonard et al.\ }{2002a}]{leon02a}
Leonard D.C., et al., 2002a, AJ, 124, 2490
\bibitem[\protect\citename{Leonard et al.\ }{2002b}]{leon02b}
Leonard D.C., et al., 2002b, PASP, 114, 35
\bibitem[\protect\citename{Li et al.\ }{2002}]{li02}
Li W., Filippenko A.V., Van Dyk S.D., Hu J., Qiu Y., Modjaz M., Leonard D.C., 2002
PASP, 114, 403
\bibitem[\protect\citename{Litvinova \& Nadyozhin\ }{1985}]{LN85}
Litvinova I.Y., Nadyozhin D.K., 1985, Sov. Astron. Lett., 11, 145
\bibitem[\protect\citename{Mazzali et al.\ }{2002}]{mazz02}
Mazzali P., et al., 2002, ApJ, 572, L161
\bibitem[\protect\citename{McGaugh \ }1991]{mcg91}
McGaugh S.S., 1991, ApJ 380, 140
\bibitem[\protect\citename{Meynet et al. }{1994}]{mey94}
Meynet G., Maeder A., Schaller G. Schaerer D., Charbonnel C., 1994, A\&AS, 103, 97
\bibitem[\protect\citename{Nadyozhin \ }2003]{nad03} Nadyozhin D.K.,
2003, to appear in Proc. of the ESO/MPA/MPE Workshop ``From Twilight
to Highlight: The Physics of Supernovae'', eds. W. Hillebrandt,
B. Leibundgut (Springer-Verlag), in press
\bibitem[\protect\citename{Pastorello et al.\ }{2002}]{pass02}
Pastorello A., et al., 2002, MNRAS, 333, 27
\bibitem[\protect\citename{Pilyugin }{2002}]{pil02}
Pilyugin L.S., 2001, A\&A, 369, 584
\bibitem[\protect\citename{Pols et al.\ }{1995}]{pols95}
Pols O.R., Tout C.A., Eggleton P.P., Han Z., 1995, MNRAS 274, 964
\bibitem[\protect\citename{Pols et al.\ }{1997}]{pols97}
Pols O.R., Tout C.A., Schr\"{o}der K.-P., Eggleton P.P., Manners J.,
1997, MNRAS, 289, 869
\bibitem[\protect\citename{Pooley et al.\ }{2002}]{pool2002}
Pooley D., et al., 2002, ApJ, 572, 932
\bibitem[\protect\citename{Popov }{1993}]{pop93}
Popov D.V., 1993, ApJ, 414, 712
\bibitem[\protect\citename{Roy \& Walsh\ }{1997}]{roy97}
Roy J.-R., Walsh J.R., 1997, MNRAS 288, 715
\bibitem[\protect\citename{Ryder et al.\ }{1993}]{ryder93}
Ryder S., Staveley-Smith L., Dopita M., Petre R., Colbert E., 
Malin D., Schlegel E., 1993, ApJ, 416, 167
\bibitem[\protect\citename{Schaller et al.\ }{1992}]{sch92}
Schaller G., Schaerer D., Meynet G., Maeder A., 1992, A\&AS, 96, 269
\bibitem[\protect\citename{Schlegel }{2001}]{schlegel2001}
Schlegel E.M., 2001, ApJ, 556, L25
\bibitem[\protect\citename{Silbermann et al.\ }{1999}]{silb99}
Silbermann N.A., et al., 1999, ApJ, 515, 1
\bibitem[\protect\citename{Smartt et al.\ }{2001a}]{sma01a}
Smartt S.J., Gilmore G.F., Trentham N., Tout C.A., Frayn C.M., 2001a, ApJ, 
556, L29 
\bibitem[\protect\citename{Smartt et al.\ }{2001b}]{sma01b}
Smartt S.J., Crowther P.A., Dufton P.L., Lennon D.J., Kudritzki R.P., 
Herrero A., McCarthy J., Bresolin F., 2001b, MNRAS, 325, 257
\bibitem[\protect\citename{Smartt et al.\ }{2001c}]{sma01c}
Smartt S.J., Kilkenny D., Meikle W.P.S., 2001c, IAUC7704
\bibitem[\protect\citename{Smartt et al.\ }{2002a}]{sma02a}
Smartt S.J., Gilmore G.F., C.A. Tout, S. Hodgkin, 2002a, ApJ, 565, 1089
\bibitem[\protect\citename{Smartt et al.\ }{2002b}]{sma02b}
Smartt S.J., Vreeswijk P. M., Ramirez-Ruiz E., Gilmore G. F.,
Meikle W. P. S., Ferguson A. M. N., Knapen J. H., 2002b, ApJ, 572, L147
\bibitem[\protect\citename{Thompson }{1982}]{thom82}
Thompson L.A., 1982, ApJ, 257, L63
\bibitem[\protect\citename{Trundle et al.\ }{2002}]{tru02}
Trundle C.A., Dufton P.L., Lennon D.J., Smartt S.J., Urbaneja M., 2002, 
A\&A, 395, 519
\bibitem[\protect\citename{Turatto et al.\ }{2003}]{turatto03} 
Turatto M., Benetti S., Cappellaro E., 2003, to appear in Proc. of the
ESO/MPA/MPE Workshop ``From Twilight to Highlight: The Physics of
Supernovae'', eds. W. Hillebrandt, B. Leibundgut (Springer-Verlag), in
press, (astro-ph/0211219)
\bibitem[\protect\citename{van den Bergh et al.\ }{2002}]{vanden02} 
van den Bergh S., Li W., Filippenko A., 2002, PASP, 114, 820
\bibitem[\protect\citename{Van Dyk et al.\ }{1996}]{vandyk96} 
Van Dyk S.D., Hamuy M., Fillipenko A.V., 1996, AJ, 111, 2017
\bibitem[\protect\citename{Van Dyk et al.\ }{2000}]{vandyk00} 
Van Dyk S.D., Peng C Y., King J.Y., Fillipenko A.V., Treffers R.R.,
Li W., 2000, PASP, 112, 1532
\bibitem[\protect\citename{Van Dyk et al.\ }{2003a}]{vandyk03a} 
Van Dyk S.D., Li W., Fillipenko A.V., 2003a, PASP, 115, 1
\bibitem[\protect\citename{Van Dyk et al.\ }{2003b}]{vandyk03b} 
Van Dyk S.D., Li W., Fillipenko A.V., 2003b, PASP, in press, astro-ph/0301346
\bibitem[\protect\citename{Vila-Costas \& Edmunds }{1992}]{vce92}
Vila-Costas M.B., Edmunds M.G., 1992, MNRAS, 259, 121
\bibitem[\protect\citename{Walborn et al.\ }{1989}]{wal89}
Walborn N. et al., 1989, A\&A, 219, 229
\bibitem[\protect\citename{Wang et al.\ }{2001}]{wang01}
Wang L., Baade D., Fransson C., Hoeflich P., Lundqvist P.,
 Wheeler J.C., 2001, IAU Circ. 7704
\bibitem[\protect\citename{White \& Malin }{1987}]{white87}
White G.L., Malin D.F., 1987, Nat., 327, 36
\bibitem[\protect\citename{Whitmore et al.\ }{1999}]{whit99}
Whitmore  B. et al., 1999, PASP, 111, 1559
\bibitem[\protect\citename{Woolley \& Stibbs\ }{1953}]{woolley1953}
Woolley R., Stibbs D. W. N., 1953, The Outer Layers of a Star,
             Clarendon Press, Oxford
\bibitem[\protect\citename{Woosley et al.\ }{1994}]{woolsey94}
Woosley S.E., Eastman R.G., Weaver T.A., Pinto P.A., 1994, ApJ, 429, 300
\bibitem[\protect\citename{Woosley \& Weaver }{1986}]{ww86}
Woosley S.E, Weaver T.A., 1986, ARA\&A, 24, 205 

\end{thebibliography}
\end{document}